# Industry Characteristics and Financial Risk Spillovers

Wan-Chien Chiu[a], Juan Ignacio Peña[a], and Chih-Wei Wang [a*]

**Abstract**

This paper proposes a new measure of tail risk spillover, namely the conditional coexceedance which is the number of joint occurrences of extreme negative returns in an industry conditional on an extreme negative return in the financial sector. The empirical application provides evidence of significant volatility and tail risk spillovers from the financial sector to many real economy sectors in the U.S. economy in the period from 2001 to 2011. These spillovers increase in crisis periods. The conditional coexceedance in a given sector is positively related to its amount of debt financing, and negatively related to its relative valuation and investment. Therefore real economy sectors which require substantial external financing, and whose value and investment activity are relatively lower are prime candidates for depreciation in the wake of crisis in the financial sector. We also find some evidence suggesting that the higher the industry's degree of competition the stronger the tail risk spillover from the financial sector.

**Keywords**: volatility spillover, tail risk, conditional coexceedance, debt financing, relative valuation, relative investment

JEL classification codes: G00 G01 G18 C33 C36

[a] Universidad Carlos III de Madrid, Department of Business Administration, c/ Madrid 126, 28903 Getafe (Madrid, Spain).

*Corresponding author

E-mail addresses: wchiu@emp.uc3m.es; ypenya@eco.uc3m.es; chwang@emp.uc3m.es.

# Industry characteristics and financial risk spillovers

**Abstract**

This paper proposes a new measure of tail risk spillover: the conditional coexceedance (CCX) which is the number of joint occurrences of extreme negative returns in an industry conditional on an extreme negative return in the financial sector. The empirical application provides evidence of significant volatility and tail risk spillovers from the financial sector to many real sectors in the U.S. economy from 2001 to 2011. These spillovers increase in crisis periods. The conditional coexceedance in a given sector is positively related to its amount of debt financing, and negatively related to its valuation and investment. Therefore real economy sectors which require relatively high debt financing, and whose value and investment activity are relatively lower are prime candidates for stock price volatility and depreciation in the wake of crisis in the financial sector. We also find some evidence suggesting that the higher the industry's degree of competition the stronger the tail risk spillover from the financial sector.

1. **Introduction**

In many countries the financial sector is a key funding source for industrial or services (i.e. real economy) firms who have limited internal funds. Therefore real economy firms' risk and return should be strongly affected by the vagaries of the financial sector and in particular to its profitability and stability. The recent events of 2007-2009 illustrate a situation where an acute distress in the financial sector implied a severe credit crunch with devastating effects on the real economy. So it is not surprising to find that the links between the financial sector and real economy sectors have been widely explored. Previous literature has investigated this linkage focusing on industrial real output (see Rajan and Zingales, 1998),[1] stock market returns (Baur, 2011),[2] and the links between other measures of returns and profitability. However the linkage between the financial sector's risk and real economy sector's risk has received little attention so far. This is surprising giving the aforementioned evidence provided by the recent financial crisis of 2007-2009. This paper aims to fill this gap in the literature presenting two contributions. First a new measure of tail risk spillovers and, second some empirical evidence on this important subject.

We explore the extent to which financial industry's risk spills over to industrial and services sectors' risk from several perspectives. First, we ask whether risk spillover from the financial sector to real economy sectors exists over the past decade and to what extent this spillover is affected by the intense distress of the financial sector in the 2007-2009 financial crises. Second, we consider both the volatility and the tail risk spillover since they provide different insights on risks. Notice that the volatility characterizes dispersion from average returns, while the tail risk concentrates on the left tail of the return's distribution. Third, we ask whether tail risk spillover is affected by the real economy sector's product market structure (competition versus concentration). Fourth, we ask whether the tail risk spillover is driven by three industry characteristics: the net debt financing, the valuation and the investment. These characteristics are closely related to the industry's investment opportunities and future perspectives.

---

[1] Rajan and Zingales (1998) point out that the industry's growth is related its dependence on external sources of finance, stemming from industry-specific technological factors which affects initial project scale, gestation period, cash-harvest period, and needs for further investments. Accordingly, there is a large literature testing the impacts of banks on the real economy at country-, industry-, and firm-level across countries and over time (Beck et al., 2000; Cetorelli and Gambera, 2001; Vives, 2001; Beck and Levine, 2002; Hoggarth et al., 2002; Beck and Levine, 2004; Claessens and Laeven, 2005; Levine, 2005; Kroszner et al., 2007; Cole et al., 2008; Dell'Ariccia et al., 2008; Chava and Purnanandam, 2011).

[2] Baur (2011) finds that the 2007-2009 crisis led to an increased co-movement of returns among financial sector's stocks across countries and between financial sector's stock returns and real economy sector's stock returns.

Furthermore we develop a new proxy for capturing financial tail risk spillover, called conditional coexceedance (CCX). The CCX measures the frequency of simultaneous extreme negative stock returns in the financial sector and in the real economy sectors. We also compute probabilities of tail risk spillover at industry-level over time distinguishing between crisis and non-crisis periods. Finally we study the determinants of the CCX measure in terms of the industry's structural characteristics. We use U.S. stock market data covering the period from 2001 through 2011. The main empirical results are (1) increases in financial industry's volatility and tail risk cause corresponding increases in the real economy sector's risk variables and the effect of this spillover is stronger in the financial crisis period; (2) The tail risk spillover from the financial sector to the real economy sector, as measured by the CCX, is stronger if the real economy industry is more competitive, uses a high proportion of net debts, and has a relatively low level of valuation and investments.

The study is related to several strands of literature. Our results are consistent with Diebold and Yilmaz (2012) in the sense that financial sector's volatility generally increases sharply and spills over to other economic sectors in times of financial distress.[3] Furthermore, our paper is related with the literature on tail risk dependence (such as Bae et al., 2003) by introducing the CCX measure and documenting role of the financial sector on the industrial sector's tail risk. We document a risk increase in real economy sectors stemming from increases in the instability of the financial sector and the consequent negative impact on the economy in agreement with Kroszner et al.

---

[3] The leading role of the financial sector with respect to aggregate volatility is also documented in Houston and Stiroh (2006), Wang (2010), and Cheong et al. (2011).

(2007). Moreover while recent evidence supports the view that the intensity of competition in a given industry has significant implications for firms' cash flows and stock returns (Hou and Robinson, 2006; Hoberg and Phillips, 2010), the significant effect of the degree of competition on the linkage of tail risks between real economy sectors and the financial sector is a novel result. Finally we provide a possible reason explaining the higher risks of highly competitive industries,[4] namely its tail risk connection with the financial sector.

Summing up, our contributions to the literature are as follows. First, we develop novel empirical methodologies for testing the effect of volatility and tail risk spillover coming from the financial sector to the real economy sectors, before and during the recent financial crisis. Second, our empirical analysis in the U.S. market over the prior decade documents that risk spillovers (both in volatility and tail risk dependence) indeed increased during the crisis period. Third, we relate the financial risk spillover measure CCX to the real economy industry's product market structure, investment opportunities and valuation. Finally we empirically document that the effect of the real economy industry characteristics on tail risk spillover measures exhibits variation across industries: it is stronger for industries that face more competitive product markets, use higher net debts, and have lower level of valuation and investment.

The study provides two key implications. First, our findings remark the close connection between the financial and real economy sectors. This result is important for practitioners because it supports the view that difficulties in the financial sector are, sooner or later, followed by large increases in uncertainty in other industries. Second, our CCX-based results imply that financial sector's extreme returns are concurrent indicator of real economy sectors' extreme returns. Therefore our CCX measure can provide warning signals of impending turmoil in stock prices of real economy firms when a financial crisis materializes. Our results may also be useful to policy makers and regulators because they are interested in evaluating the possible economic costs due to financial crises. In particular, one of the key objectives of the Basel III capital accord is to reduce the risk of spillover from the financial sector to the real economy. Given that the adoption of Basel III norms is expected to reduce the frequency and severity of these spillovers our modeling approach can be used to assess the extent to which Basel III actually performs as expected. Finally the fact that real sectors which require debts, and whose value and investment activity are relatively lower are prime candidates for depreciation in the wake of crisis in the financial sector, has obvious implications for investors.

---

[4] See Valta (2012).

The remainder of the paper proceeds as follows. The next section discusses how risk spillover and industry characteristics interact. Section 3 addresses the empirical methodologies. Section 4 introduces the database and the construction of the industry characteristic variables. Section 5 discusses the empirical results. Robustness test are provided in Section 6. Section 7 concludes.

## 2. Risk spillover and industry characteristics

### 2.1. *Volatility spillover*

The evidence on the extent to which risk increases in the financial sector spill over to risk measures in industrial sectors is relatively scarce. Houston and Stiroh (2006) find that in the U.S. economy, financial sector's volatility has had a significant and negative impact on economic growth in the period from 1985 through 1994.[5]

---

[5] It was a very turbulent period for the U.S. banking sector. Large banks suffered huge losses from loans to developing countries. Savings & Loans failures peaked in 1988 and 1989.

Concentrating on volatilities, Wang (2010) shows that the financial sector's volatility leads non-financial sectors' in the U.S. market in the period from 1963 to 2008 and Cheong et al. (2011) results support this view in the UK economy from 1990 to 2010. But as far as we know there is no evidence from other economic areas or time periods.

A related question is whether the recent financial crisis affects the risk spillover mechanism from the financial sector to the real economy. If a sudden loss happens within the financial system, its contractionary impact on real economy sectors is bound to be strong (see Kroszner et al., 2007). The 2007-2009 crisis has led to an increased co-movement between financial sector's stock returns and real economy's stock returns (Baur, 2011). Recent evidence reports that the 2007-2009 financial crisis has had a negative impact on industrial sectors' investment activities (Campello et al., 2010).[6] Finally the shortage of external funding weakens firms' operating flexibility during crisis periods because firms face budget constraints, reducing their investments, and then increasing their equity risk (Ortiz-Molina and Phillips, 2013). Based on the previous evidence, we hypothesize in turn that

*Hypothesis 1: There are volatility spillovers from the financial sector to real economy sectors. These spillovers are stronger during financial crisis.*

*2.2. Tail risk spillover and product market structure*

Differently from volatility which only characterizes dispersion from average returns, tail risk concentrates on the left tail of returns' distribution. In fact, the left tail risk has been addressed in the extant literature. Bae et al. (2003) study extreme co movements between stock returns using exceedance correlations. Hartmann et al. (2004)

---

[6] They report that 86% of U.S. CFOs canceled or postponed attractive investment opportunities due to the inability to borrow externally.

employ a non-parametric measure, using extreme value theory (EVT) to gauge spillover effects between stock and bond markets.

As for the linkage of tail risk between financial and industrial sectors, Christiansen and Ranaldo (2009) applies the method of Bae et al. (2003) to compare the financial integration of the old and new EU countries' stock markets, and find strong persistence of coexceedance for both regions, especially in lower tail coexceedance. Beine et al. (2010) measure stock market coexceedance using quantile regression, showing that financial liberalization leads to a rise in left tail's co movements. Fry et al. (2010) focus on the coskewness of market returns, showing spillover effects between the real estate and equity market within and between countries for the Hong Kong crisis in 1997 and in the subprime crisis in 2007. Hence, we expect tail risk spillovers to exist and be especially intense in downturns – a point which also has become very evident during the 2007-2009 financial crises. We hypothesize in turn that

*Hypothesis 2a: There are tail risk spillovers from the financial sector to real economy sectors. These spillovers are stronger during financial crisis.*

Next, we ask whether tail risk spillover is affected when industrial sectors have different product market structure. Valta (2012) studies U.S. firms from 1992 to 2007 and find strong empirical evidence that banks charge significantly higher interest rates in loans given to firms in competitive environments due to their higher default risk and lower firms' asset liquidation value. Moreover Ortiz-Molina and Phillips (2013) find that the cost of inflexible operations is higher for firms that face more competitive risk in product markets. Therefore, we guess that competitive industries may experience larger financial tail risk spillover than concentrated industries. Thus, we hypothesize as follows:

*Hypothesis 2b: The tail risk spillover from the financial sector is stronger (weaker) the higher (lower) the degree of competition in a given industry.*

*2.3. Tail risk spillover and industry characteristics*

We further investigate how industry characteristics may have different impact on the linkage between the financial sector's and real economy's risks. We consider three different industry-level variables: (1) industry net debt issuance (2) industry valuation, and (3) industry investment.

One possible channel by means of which the financial sector may affect industrial firm's growth and risks is the firm's external financing dependence. For example, starting from Rajan and Zingales (1998), scholars have paid much attention to how the degree of competition in the banking sector and how the dependency on external financing across nonfinancial industries have different impact on nonfinancial industries' growth and structure.[7] The evidence shows that sectors that are highly dependent on external finance tend to experience a substantially greater contraction of value added during a banking crisis (Kroszner et al, 2007). Real asset illiquidity increases for firms that have less access to external capital or are closer to default and face more competition (Ortiz-Molina and Phillips, 2013).

In particular, debt is the main source of external finance for firms' operating flexibility and for real investment activities (Valta, 2012). Industries that depend heavily on debt financing, sometimes encounter difficulties raising funds from the financial sector. In normal times, firms have more chances to find funding sources either from the financial sector or from selling assets. However when the financial

---

[7] Vives (2001) posits that the degree of competition in the financial sector is crucial when firms look for external financing. By adopting an industrial organization-based measure in assessing the competition of a banking sector, it has been shown that greater competition in the banking sector fosters the growth rate of financially dependent industries (Claessens and Laeven, 2005).

sector is in crisis, credit constraints may appear. It is also likely that under these circumstances asset markets are also stressed. Thus, industrial sectors that have higher level of net debt may face a more negative impact when the financial sector is distressed, leading to an increase in tail codependence. In summary, we hypothesize that

*Hypothesis 3a: The higher the industry's debt financing the higher the tail risk spillover from the financial sector.* .

Ortiz-Molina and Phillips (2013) suggest that the problem arising from illiquid asset markets is higher for firms with low valuations (low market-to-book ratios). Thus, industrial sectors with higher valuations can obtain higher prices when selling assets, reducing their dependence from the financial sector. In addition, Fama and French (1995) and Chen and Zhang (1998) show that firms with low market-to-book ratios have persistently low earnings, higher financial leverage, and more earnings uncertainty. That is, we can expect that low market-to-book (low valuation) firms experience greater distress risks, and then higher tail risks in connection with the financial sector during weak economic times. We hypothesize that,

*Hypothesis 3b: The higher the industry's valuation the lower the tail risk spillover from the financial sector.*

The level of investment on industrial sectors could also drive the financial tail risk connection. If firms cannot fully exploit their investment opportunities, they risk losing these opportunities and market share to rivals (Valta, 2012). In other words, the *ex ante* higher investment means firms fully exploit their investment opportunities, and probably have more *ex post* internal financing resources, thus reducing the dependence of tail risks stemming from distressed financial industry.

*Hypothesis 3c: The higher the industry's investment the lower the tail risk spillover from the financial sector.*

In summary, we expect that some industry characteristics affect tail risk spillover:

It should be stronger for more competitive industries, for industries which bear higher debts, and for industries which have low level of valuation and investment.

## 3. Empirical methodologies

### 3.1. Volatility spillovers

To model volatility spillovers we follow Liu and Pan (1997), and employ a two-stage VAR-GARCH approach to model the mechanism of volatility transmission. However we modify their approach both in the first and second stage to suit the problem at hand. Specifically, in the first stage, we model two equity index return series, corresponding to the U.S. financial sector and to a given U.S. industrial sector respectively including both series in a VAR system. In doing so, we adjust for autocorrelations in each series as well as for cross correlations between series. The residuals obtained after fitting the VAR model,[8] are denoted $r_{i,t}$ and $r_{FIN,t}$, where $i$ and $FIN$ stand for any non-financial industrial sector and for the financial sector respectively. Next, we standardize the series $r_{FIN,t}$ by means of a GARCH(1,1) process as follows:

$$r_{FIN,t} \sim N\left(0, \sigma^2_{FIN,t}\right) \qquad (1)$$

$$\sigma^2_{FIN,t} = \omega_{FIN} + \alpha_{FIN} r^2_{FIN,t-1} + \beta_{FIN} \sigma^2_{FIN,t-1} \qquad (2)$$

$$\frac{r_{FIN,t}}{\sigma_{FIN,t}} = e_{FIN,t} \sim N(0,1) \qquad (3)$$

where the standardized series is $e_{FIN,t}$. In the second stage, we model volatility spillovers using:

---

[8] The optimal lag in the VAR system was chosen using the Schwarz-Bayesian (BIC) criterion.

$$r_{i,t} \sim N(0, \sigma_{i,t}^2) \tag{4}$$

$$\sigma_{i,t}^2 = \omega_i + \alpha_i r_{i,t-1}^2 + \beta_i \sigma_{i,t-1}^2 + \gamma_{i1} e_{FIN,t-1}^2 + \gamma_{i2} e_{FIN_{crisis},t-1}^2 \tag{5}$$

Where $e_{FIN_{crisis},t-1}^2$ equals $e_{FIN,t-1}^2$ during crisis periods and equals zero otherwise.[9] There is a volatility spillover from the financial sector to the industry $i$ if the coefficient $\gamma_{i1}$ in (5) is significantly positive. If the financial crisis amplifies the spillover we expect the coefficient $\gamma_{i2}$ in (5) to be significantly positive as well.[10] As a robustness test we also relax the Normality assumption in (1) and in (4) by allowing residuals to follow Student t-distribution.[11]

*3.2. Tail risk spillover*

*3.2.1. Conditional coexceedance CCX*

We define an extreme return, or exceedance, as one that lies either below (above) the αth (1-αth) quantile of the marginal return distribution. Formally an exceedance for industry $i$ at time $t$ is defined as

$$I_t^i(c) = \begin{cases} 1, & \text{if } r_t^i \in c \\ 0, & \text{otherwise} \end{cases}, \quad t = 1,...,T \tag{6}$$

---

[9] To obtain a crisis period encompassing all major financial and economic events, we adopt timelines provided by the Bank for International Settlements (BIS, 2009). The BIS study separates the timeline in four phases from the third quarter in 2007 until the end of 2009. Phase 1 spans from Q3 in 2007 until mid-September 2008 and is described as ''initial financial turmoil''. Phase 2 (mid-September 2008 until late 2008) is defined as ''sharp financial market deterioration'', phase 3 is a phase of macroeconomic deterioration (until Q1 2009). Hence, in this paper, we define the crisis period starting in the second quarter of 2007 and ending in the first quarter of 2009.

[10] In the equation (5), the parameter $\gamma_{i1}$ measures the spillover of shocks from financial sector to another sector in normal periods; the parameter $\gamma_{i2}$ measures the additional contribution of the crisis period to this spillover, and thus the ($\gamma_{i1} + \gamma_{i2}$) reflects total effect of volatility spillover during crisis periods. If $\gamma_{i2}$ is positive (negative), there is an additional increased (decreased) transmission of unexpected shocks from US financial sector to another sector i in the crisis period compared to normal periods. A similar setting is applied in Baur (2011, JBF), where the author aims to examine whether there is a contagion from the financial sector to the real economy by looking at equity returns during 2007-2009.

[11] The results remain do not materially change in comparison with the Normality assumption. Detailed results are available on request.

where $I_t^i(c)$ is the indicator function that equals one when the return $r_t^i$ belongs to the set $c$ and equals zero otherwise. We arbitrarily define $c$ as the set of asset returns located below the 5$^{th}$ quantile of the marginal distribution of daily returns.[12] Next we posit a new measure of tail risk spillover from the financial sector to non-financial sectors in the spirit of Bae et al. (2003) by concentrating on the occurrence of simultaneous negative extreme returns across assets as the key element of spillover.[13] We define our new measure, named "*conditional coexceedance*" (henceforth CCX) for a non-financial industry $i$ at time $t$ as

$$CCX_t^i = I_t^i(c) \times I_t^{FIN}(c), \ t = 1,...,T \qquad (7)$$

where $CCX_t^i$ equals one if a non-financial industry $i$ and the financial sector both have exceedances at time $t$. $CCX_t^i$ is equal to zero otherwise. We name the measure "*conditional coexceedance*" in order to stress the key role of the financial sector in our measure and to distinguish it from the unconditional measures used in Bae et al. (2003).[14] The intuition supporting this measure is that a non-financial industry is exposed to tail risks which are dependent on the simultaneous occurrence of extreme negative returns in the financial sector.

With this measure we can compute the observed frequencies (likelihoods) of CCX for a given non-financial industry $i$. To do so, we compute the proportion of CCX over a fixed time horizon as follows,

$$\text{Prob}^i = n_1^i / (n_1^i + n_0^i) \qquad (8)$$

---

[12] We use the 5% quantile as the baseline. We also employ the 1% and 2.5% quantiles as a robustness test, and the results are similar to those based on 5% quantile. Detailed results are available on request.
[13] Bae et al. (2003) claim that coexceedances are superior to the correlation coefficient if non-linearities in market behavior exist since coexceedances are not restricted to describe linear market behavior.
[14] Bae et al. (2003) define a coexceedance of $n$ at $t$ as the situation when $n$ assets present exceedances on the same day $t$, whereas in our setting, one of units in a coexceedance must be the financial sector.

where $n_1^i$ is the number of 1's and $n_0^i$ is the number of 0's in the indicator series in equation (7). Furthermore we separate $\text{Prob}^i$ into two components to identify the relative frequency of CCX during crisis and non-crisis periods for an industrial sector *i*, as follows,

$$\text{Prob}^i_{crisis} = \frac{n^i_{1,crisis}}{\left(n^i_{1,crisis} + n^i_{0,crisis}\right)}, \text{ and } \text{Prob}^i_{non-crisis} = \frac{n^i_{1,non-crisis}}{\left(n^i_{1,non-crisis} + n^i_{0,non-crisis}\right)} \quad (9)$$

where $n^i_{1,crisis}$ ($n^i_{0,crisis}$) is the number of 1's (0's) in the indicator series during the crisis period, and $n^i_{1,non-crisis}$ ($n^i_{0,non-crisis}$) is the number of 1's (0's) in the indicator series during the non-crisis period.[15]

*3.3. Determinants of CCX*

Given that CCX take only non-negative integer values, we use the Poisson panel regression model to study their possible determinants. The dependent variable is the total number of daily conditional coexceedances observed in one quarter in a given industry, and the explanatory variables are proxies for industry debt financing, industry valuation, industry investment, and control variables. We choose these explanatory variables because the evidence in Hoberg and Phillips (2010) supports their prominent role in identifying the conditions that likely surround industry booms and busts.

Given that some of our explanatory variables are generated regressors, and therefore prone to the problem of errors-in-variables, standard OLS-based methods are suboptimal. Instead, we apply the estimation method based on the Generalized Method of Moments (GMM) with suitable instrumental variables. This approach improves the efficiency and consistency of the estimates and mitigates to a considerable extent the errors-in-variables problem.[16] The baseline panel model specification is as follows:

$$CCX_{i,t} = EXP\left\{\alpha + \sum_{n=1}^{N}\beta_n X_{n,i,t-1} + \sum_{l=1}^{L}\delta_l control_{l,i,t-1} + Industry_{dummy} + Time_{dummy}\right\} + \varepsilon_{i,t}$$

(10a)

$$CCX_{i,t} = EXP\left\{\begin{array}{l}\alpha + \sum_{n=1}^{N}\gamma_n X_{n,i,t-1} \times D_{crisis,t-1} + \sum_{n=1}^{N}\omega_n X_{n,i,t-1} \times D_{non-crisis,t-1} \\ + \sum_{l=1}^{L}\delta_l control_{l,i,t-1} + Industry_{dummy} + Time_{dummy}\end{array}\right\} + \varepsilon_{i,t} \quad (10b)$$

where the dependent variable $CCX_{i,t}$ is the actual number of coexceedances observed quarterly for industry $i$. The dependent variable is regressed on the set of one-quarter lagged explanatory variables $X$ and *control*. The vector of variables $X_{n,i,t}$ contains the industrial characteristic variables for industry $i$ (net debt financing, spread from a normative value, and spread from a normative investment ). The vectors $D_{crisis,t}$ and $D_{non-crisis,t}$ are dummy variables for the crisis period and the non-crisis period. Both variables are designed to test whether the relationship between CCX and industry effects is sensitive to the emergence of financial crisis. In particular, we define the crisis period is between the third quarter of 2007 and the first quarter of 2009. The vector of $control_{l,i,t}$ contains variables related to other industries' characteristics: volatility of profitability, debt cost, earnings per share, and size. We expected the impact of the volatility of profitability on CCX to be positive. The reason is that industries with higher profit instability would probably need more external financing and therefore are more exposed to financial sector upheavals. A similar reasoning applies to firms with relatively high cost of financing. On the other hand we expect big firms and firms with high earnings per share to be relatively less exposed to financial sector's disturbances. Finally, the model also includes industry and time dummies.

---

[15] When testing for equality of the likelihood during crisis and during non-crisis period, the standard Normality-based tests are inappropriate because, by construction, both variables follow non-Gaussian distributions. Therefore we employ the Wilcoxon rank-sum test which is a nonparametric alternative to the standard two sample t-test. The Wilcoxon test is based solely on the order in which the observations from the two samples. Given that we want to know whether the distribution of Prob$_{crisis}$ is shifted to the right of distribution Prob$_{non-crisis}$, in the empirical results section we report tests for this one-sided alternative.

---

[16] We also employed the conditional maximum likelihood estimation (CMLE) method in the spirit of Silva and Tenreyro (2006). Although most estimated coefficients are largely in agreement under both methods, a few of them are not similar. These facts suggest that the errors in variables problem may have material influence on the results and therefore we choose to present the GMM estimations in the main text. The CMLE results are available on request.

## 4. Database and industry characteristic variables

*4.1. Database: Sample selection and industry classification*

The empirical results are based on stock prices of firms traded in the U.S. market with available data in the Center for Research in Security Prices (CRSP) during the period from January 2001 to December 2011.[17] We aggregate firm-level returns into value-weighted industry-level returns, in order to test the industry-level volatility spillovers and tail risk spillovers. We work with 73 non-financial industries and one financial sector [18] (see Appendix A) by adopting the three-digit NAICS codes. Regarding the construction of industry characteristic variables, we use quarterly accounting information obtained from COMPUSTAT databases.

We use the fitted Herfindahl-Hirschman index (HHI) proposed by Hoberg and Phillips (2010) to identify competitive and concentrated industries.[19] Specifically we define competition (concentration) if the fitted HHI is in the lowest (highest) quartile of the yearly sample distribution (see Valta, 2012).[20]

*4.2. Industry characteristic variables*

We construct three industry-level proxies for new opportunities and future prospects, including (1) net debt financing; (2) spread between the actual value and a normative value; (3) spread between the actual investment and a normative investment. The detailed description of these variables follows.

---

[17] The starting year of 2001 is chosen because Bureau of Labor Statistics data (only available from 2001) is required in order to classify a sector's degree of competition.
[18] The three-digit NAICS code for firms belonging to the financial sector is 521-525 and 531-533.
[19] The details on the fitted HHI are in Appendix B.
[20] We also use the upper 10% and lower 10% of fitted HHI as a robustness, and the results do not change materially.

*(i)    Net Debt Financing*

For a given firm, we measure its net debt financing in a given quarter as the firm's net debt issuing activity divided by total assets. The firm's net debt issuing activity is equal to long term debt issuance minus long term debt reduction (Hoberg and Phillips, 2010). Consequently, the industry's net debt financing is the total amount of net debt financing for all firms in the industry divided by total industry assets.

Those industries demanding higher levels of debt-like financing should be more vulnerable to financial sector's distress. Hence, we expect the tail risk of industries with a higher dependency on debt financing to be strongly linked with financial sector's tail risk which implies a positive coefficient in the regression.

*(ii)    Spread between the Actual Value and a Normative Value*

We define an industry "spread" time-series valuation (hereafter, spread valuation) using the procedure proposed in Hoberg and Phillips (2010). First, we compute the normative value based on the valuation model in Pastor and Veronesi (2003). We regress the log of the market-to-book ratio, $\log(M/B)$ of the reciprocal of one plus firm age (AGE), a dividend dummy (DD), firm leverage (LEV), the log of total assets (SIZE), current firm ROE, and the volatility of profitability (VOLP) for each firm $i$.

$$\log\left(\frac{M}{B}\right)_{i,\tau} = a + bAGE_{i,\tau} + cDD_{i,\tau} + dLEV_{i,\tau} + e\log(SIZE_{i,\tau}) + fVOLP_{i,\tau} + gROE_{i,\tau}$$

$$, \tau = t-12,\ldots,t-1. \quad (11)$$

where book equity is constructed as stockholders' equity plus balance sheet deferred taxes and investment tax credit minus the book value of preferred stock. The market equity is computed by multiplying the common stock price by common shares outstanding. LEV is total long-term debt divided by total asset. ROE is earnings divided by last year's book equity. Earnings are calculated as income before extraordinary items available to common stockholders plus deferred taxes from the income statement, plus investment tax credit. The VOLP is calculated by regressing the ROE on lagged ROE for all firms in each industry and taking the variance of the residuals. We winsorize VOLP and ROE variables at both the 1% and the 99% level. We estimate the valuation regression above using a rolling 12-quarter window of lagged data in each industry to get a set of coefficients that we apply to each quarter t to get a measure of predicted valuation. Finally we compute spread (unpredicted) valuations, which we henceforth call spread valuation, expressed as follows:

$$\text{Spread Valuation}_{i,t} = \log\left(\frac{M}{B}\right)_{i,t} - \text{Predicted}\left(\log\left(\frac{M}{B}\right)_{i,t}\right)$$

(12)

We winsorize this measure at the 1% and 99% level. The spread valuation determines a firm's intrinsic worth based on its estimated future free cash flows. The industry-level spread valuation is the average over all firms in each industry. We posit that the higher the level of spread valuation, the lower its funding dependence on the financial sector. The reason is that highly valued industries usually have fewer problems in selling their assets if needed, and therefore are less dependent on external financing. The implication is that we should expect a negative regression coefficient.

*(iii)   Spread between the Actual Investment and a Normative Investment*

Similar to the case of spread valuation, we define an industry "spread" time-series investment (hereafter, spread investment) as the variable of spread between the actual investment and a normative investment. First, we compute a normative investment based on the proposed methodology in Hoberg and Phillips (2010). That is, we regress the log of capital expenditures divided by lagged property, plant, and equipment, TOBINQ is Tobin's *q,* and the other independent variables are the same as the ones in the regression (11),

$$\log\left(\frac{Invest_{i,t}}{PPE_{i,t-1}}\right) = a + bTOBINQ_{i,t} + cDD_{i,t} + dLEV_{i,t} + e\log\left(SIZE_{i,t}\right) + fVOLP_{i,t} + gROE_{i,t} \quad (13)$$

From the above model, we calculate spread investment as the difference between the actual investment and the predicted investment as follows

$$\text{Spread Investment}_{i,t} = \log\left(\frac{Invest_{i,t}}{PPE_{i,t-1}}\right) - \text{Predicted}\left(\log\left(\frac{Invest_{i,t}}{PPE_{i,t-1}}\right)\right) \quad (14)$$

We measure the industry-level spread investment by averaging over all firms in each industry. The *ex ante* higher spread investment means that firms fully exploit their investment opportunities, and probably have more *ex post* internal financing resources, thus reducing their dependence from the financial industry. The previous reasoning implies a negative regression coefficient.

## 5. Empirical results

*5.1.1. Summary statistics*

Descriptive statistics of industry returns during the full sample period and during the financial crisis are presented in Table 1 and Table 2 respectively. The number of firms within each sector varies considerably. For instance, Computer and Electronic Product manufacturing is the most populated sector and contains an average of 627 firms. On the other hand, Wholesale Electronic Markets and Agents and Brokers only contain an average of 2.2 firms. Average returns are usually positive or zero in the full sample and, as expected, they are usually lower in the crisis period than in the full sample, excepting two industries, Crop Production, and Educational Services. Regarding the standard deviation, the crisis sample presents larger figures compared with the full sample, as expected. The least volatile sectors, both in the full sample (first figure) and in the crisis period (second figure) are Food Manufacturing (0.010, 0.016), Beverage and Tobacco (0.011, 0.015) and Chemical Manufacturing (0.012, 0.017). In the full sample the most volatile sectors are Hospitals (0.028) and Nursing & Residential Care (0.028). The two same sectors which are the most volatile in the full sample are also the most volatile in the crisis period with daily volatilities of 0.045 and 0.053 respectively. Interestingly, they present huge one day negative returns; specifically Hospitals -25.1% (2008/6/12) and Nursing -23.5% (2008/10/27). The reasons behind these huge drops in prices are as follows. The Hospitals sector contained only two stocks that day and one of them (Chindex International, Inc), announced an unexpectedly bad result for its 2008 income, dragging the sector index with it. The Nursing sector also contained only two stocks and one of them (Sunrise Senior Living Inc.) suffered heavy losses that day due to negative news about its expected profitability. The Air Transportation also exhibits a remarkable negative return (-31.1%) in one day (2001/9/17). The explanation of this huge decrease is because that day the trading of airline stocks resumed after the events of 9/11 and investors dumped these stocks given the uncertain prospects of the airline industry after the 9/11 attacks.

It is worth mentioning that the Financial sector has an average volatility of 0.019 in the full sample and of 0.035 in the crisis sample, presenting lower volatility than the average of all sectors in the full sample (0.0198) and higher volatility than the average of all sectors in the crisis period (0.030). Jarque-Bera statistics are highly significant for both sample periods suggesting that the normality assumption is unlikely to hold.

[INSERT TABLE 1 HERE]

[INSERT TABLE 2 HERE]

Table 3 reports descriptive statistics of the dependent and explanatory variables in the Poisson regression analysis (see Eq. (10)). The dependent variable, CCX, obviously becomes larger of crisis period than it of full sample. The variable net debt financing (ND_I) is always positive on average implying that real economy firms have positive net debt issuing activity both in the whole period and in the crisis sub period. Note however that debt financing needs increased in this last sub period. The spread valuation (VAL_I) is very small and positive in the full sample (0.1%) but strongly negative during the crisis with an average undervaluation close to 17%. Regarding to the spread investment (INV_I), the actual investment is always higher than the predicted investment and much higher (9.9%) during the crisis than in the full sample (4.6%) on average. It is surprising since investments should be constricted as financial sector in distress, but its median value is slightly lower during the crisis and its standard deviation is much higher (1.59) during the crisis compared to it (0.95) over the whole sample. The volatility of profits, leverage and debt costs all increase in the crisis period, pointing out the overall increase in firms' riskiness. Surprisingly, net income is always negative (-0.2%, -0.7%), implying negative earnings per share in the full period as well as during the crisis. Most variables present noticeable increases in their volatilities during the crisis, as expected.

[INSERT TABLE 3 HERE]

*5.1.2.  Volatility spillovers*

Table 4 displays the results of the volatility spillover model shown in equations (1)-(5). The table 4 shows the coefficient estimates in the full sample ($\gamma 1$) and the additional impact due to the crisis period ($\gamma 2$) with their corresponding t-statistics. The notation "c" stands for the existence of spillover. Several important findings are worth noting.

First, the coefficient estimate reflecting spillovers in the full sample ($\gamma 1$) shows that 55 out of 73 industries suffer significant volatility spillovers from the financial sector (around 75% of all cases). The coefficient varies significantly across the sample of sectors. For example, the average size of the $\gamma 1$ coefficient is 0.34, but there are some industries with particularly strong spillover effects, such as Air Transport (2.35), Repair and Maintenance (0.89) and Amusement, Gambling, and Recreation Industries (0.81). On the other hand, a number of industries do not seem to be affected by volatility spillovers, namely Textile Product Mills, Fabricated Metal Product Manufacturing, Furniture, and Nursing among others.

Second, the coefficient estimate of $\gamma 2$ measures the additional crisis-specific influence. A positive and statistically of $\gamma 2$ implies additional volatility spillovers in crisis periods. Table 4 shows that 61 out of 73 industries (84% of all cases) exhibit an increased volatility spillovers originating from the financial sector. Compared to the situation in normal times (55 out of 73), this spillover effect is stronger in bad times. For those 55 industries that suffer volatility spillover in normal times, 48 industries (87%) have additional volatility exposure to the financial sector during the crisis period. On the other hand, for those 18 industries that are not exposed to volatility spillovers from the financial sector in normal times, 13 industries of them (72%) are now exposed in bad times. This result implies that industries more sensitive to movements of financial sectors' volatilities in normal times are also more likely to have additional spillover effects during crises. The average size of the $\gamma 2$ coefficient is 0.55, but the coefficient varies across sectors. This coefficient estimate is above 1.26 in three industries: Air Transport (2.38), Hospitals (1.26), and Ambulatory Health Care Services (1.26). In contrast, the coefficient estimate is below 0.06 in three industries: Data Processing, Hosting, and Related Services (0.06), Animal Production and Aquaculture (0.06), and Beverage and Tobacco Product Manufacturing (0.006). Overall the evidence suggests that very few sectors are immune to the effects of the volatility spillovers in crisis times, and some sectors are more severely affected than others.

Furthermore, the degree of volatility spillover in the crisis period can be calculated by adding up the coefficient estimates γ1 and γ2. The industry of Air Transport has the highest value of (γ1+γ2) with 4.74, followed by Repair and Maintenance with 1.98. Compared to the average size of the γ1+γ2 coefficient of 0.9, the two industries' volatilities are strongly sensitive to the volatility of the financial sector.

Finally, there are only five industries whose volatilities seem to be unaffected in any circumstances by the financial sector's volatility; they are Animal Production and Aquaculture, Textile Product Mills, Electronics and Appliance Stores, Pipeline Transportation, and Personal and Laundry Services.

Summing up the overall evidence is consistent with our Hypothesis 1 in the sense that, in most cases, the empirical evidence supports the existence of volatility spillovers from the financial sector to real economy sectors and these spillovers are usually stronger in the crisis period. The results suggest that volatility shocks originating from the financial sector is a source of risks that affect most other real economy sectors.

[INSERT TABLE 4 HERE]

*5.2. Tail risk spillover*

Here we present the results for the variable CCX, our proxy for financial tail risk spillover, where the intuition is that the larger the measure the higher the spillover exposure. Table 5 shows the likelihoods of CCX over the whole sample, and during the crisis and non-crisis period. In the full sample and on average, on less than 3% of the days there are simultaneous negative extreme returns in the financial sector and in the industries. However the situation changes dramatically when we split the sample into crisis periods and non-crisis periods. In the first case, in more than 8% of the days there are simultaneous extreme negative returns, whereas in the second case this figure is less than 1.4%. Furthermore, and as the Wilcoxon test indicates, these averages are statistically different at any reasonable significance level. In fact all industries exhibit an increased tail risk co-movement with the financial sector in the financial crisis period compared with non-crisis period.[21] The industries whose tail risks are more exposed to financial sector distress in crisis situations are Fabricated Metal (12%), Printing (12%), and Forestry and Loggin (11.34%) and the less exposed are Animal Production (3.8%), and Wholesale Electronic Markets (4.1%). Overall, the empirical evidence supports Hypothesis 2a in the sense that we do find tail risk spillovers from the financial sector to real economy sectors, these spillovers being amplified during the period of financial crises.

[INSERT TABLE 5 HERE]

*5.3. Competition versus concentration*

In order to test hypothesis 2b, this section addresses whether the strength of tail risk spillover increases with the degree of industry's competitiveness. We measure the CCX at yearly frequency since the competitive identification for each industry is updated every year. Specifically, for a given year, we use daily returns to compute the CCX, and we update the sample year by year. We classify the industries' degree of competition using the fitted HHI measure. We consider an industry to be concentrated if it belongs to the upper 25% of fitted HHI and be consider it to be competitive if it belongs to the lower 25% of fitted HHI.

---

[21] As a robustness test we change the starting point of the full sample from 2001 to 2003. The results do no change materially. We also changed the starting of the crisis period to the first quarter of 2007. Results support even more our hypotheses in the sense that all industries exhibit an increased tail risk co-movement with the financial sector in the financial crisis period compared with non-crisis period. Details are available on request.

[INSERT TABLE 6 HERE]

Table 6 presents the likelihoods of CCX for competitive and concentrated industries in Column 3 and 4 respectively for each year as well as for the total sample. The CCX measure is always higher for competitive industries than for concentrated ones although the differences between them are only significant (at 5%) in two years. However the average difference using all years is positive and significant at the 1% level. Therefore the data gives some support to the hypothesis 2b, in the sense that competitive industries' tail risk may be more affected than concentrated industries' by extreme negative returns in the financial sector.

*5.4. CCX and industry characteristics*

*5.4.1. The effect of industry characteristics*

Next, we now consider the impact of the variables of industry characteristics on outcomes of CCX which represents the tail risk spillover from the financial sector to other industries. Table 7 displays the results of industry-level regressions of CCX on net debt financing, spread valuation, and spread investment, respectively (Eq. (10a) and (10b)). In the most complete model, (Model 5) we find that the impact of net debt financing is positively related to CCX, whereas spread valuation and spread investment are all negatively related to CCX, as expected. The results are consistent with our hypothesis 3a, 3b, and 3c. The variable with the highest economic impact[22] on CCX is the spread valuation followed by size. The control variables have the expected signs and they are all significant. The degree of fit as measured by the pseudo-$R^2$ increases when industry-level variables are included in the equation,[23] but the bulk of this increment is associated with spread valuation variable which confirms its relevant role.

---

[22] The economic impact of the variable X on the dependent variable Y is measured by the percentage change of Y when there is an increase in one standard deviation of X. Formally it is computed as $[\exp(b_x s_x)*Y_m - Y_m]/Y_m$ where $b_x$ is X's regression coefficient, $s_x$ is X's volatility and $Y_m$ is the average sample value of Y.

[INSERT TABLE 7 HERE]

Then we split each of the industry characteristics variables into non-crisis and crisis periods (Eq. (10b)). The results are in Table 8. The effect of the variable net debt is weakly significant in the non-crisis periods, but clearly significant in the crisis periods as well as in the total sample. The spread valuation and spread investment are significant in both periods. Thus it seems that in the crisis period the impact of the net debt variable on CCX increases, but the main results obtained in Table 7 remain unchanged giving additional support to hypothesis 3a, 3b, and 3c.

[INSERT TABLE 8 HERE]

*5.4.2. The effect of competitive and concentrated industries*

The degree of competition fundamentally affects the firms' operating decisions and the riskiness of their business environment. As such, it is important to understand whether and how this characteristic of the product market structure affects financial tail risk spillovers. This section follows the main analysis by using Eq. (10a,b) on industry characteristic variables, but we focus on competitive and concentrated industries separately.

[INSERT TABLE 9 HERE]

Comparing competitive industries (see Panel A of Table 9) with concentrated industries (see Panel B of Table 9), the expected signs remain, but there are changes in statistical significance.[24] First, in the case of competitive industries, spread investment is not significant in the full sample although it is significant in the non-crisis period. There is some evidence of a positive (negative) impact of debt financing (spread investment) on CCX in the crisis (non-crisis) period. However spread valuation and debt financing remain significant explanatory variables for CCX in all periods. Therefore for competitive industries the variables that seem to keep their explanatory power more consistently are net debt financing and spread valuation. The results are different in the case of concentrated industries. In this case the only significant variable is spread valuation. One possible reason of this non-significant effect could be related with the evidence in Ghosal and Loungani (1996) in the sense that the impact of price uncertainty on the decision of investment is small for the relatively non-competitive industries, since the outcome depends on the strategic interaction within the industry. Therefore, the level of spread investment in concentrated industries should not necessarily provide useful information in predicting CCX.[25]

---

[23] For panel regression models standard measures of fit are not well defined. We take the square of the correlation between the original and fitted values of the dependent variable as the pseudo-$R^2$.

## 6. Robustness tests

We consider a battery of robustness checks. First, we use distance-to-default as proxy for tail risks. Second, we re-examine our main analysis by discarding some industries with abnormal returns. Third, we test the reverse causality from economic sectors to financial industry. Finally, we compute a dynamic CCX computed using a three-day window and study its relationship with industry characteristics.

*6.1. The exposure of distance-to-default*

---

[24] To save space, we do not report estimation results for control variables in Table 9.
[25] It should be kept in mind that these results are based on a subsample of somewhat extreme observations (highly competitive or concentrated industries). It is worth emphasizing that we define competition (concentration) if the fitted HHI is in the lowest (highest) quartile of the yearly sample distribution.

We also adopt the aggregate Merton's distance-to-default (DD) metric to test tail risk spillovers, and to check to what extent the spillover effect from the financial sector to other industries becomes worse during the crisis period.[26] We calculate monthly DD for each firm within a given sector, first using a rolling window and then averaging all firms to build an industry-level DD. In order to match the sample frequency of our control variables (quarterly), we aggregate the monthly DD to quarterly frequency. We run the following two regressions to analyze the tail risk spillover from the financial sector to other industries and to check whether the spillover effect becomes more severe in the wake of financial crises.[27] The panel regressions are as follows:

$$DD_{i,t} = \alpha_i + \beta_{i1} \times DD_{fin,t-1} + \sum_{j=1}^{J} \beta_{ij} \times control_{j,t-1} + Industry_{dummy} + Time_{dummy} + \varepsilon_{i,t} \quad (15)$$

$$DD_{i,t} = \alpha_i + \beta_{i1} \times DD_{fin-non-crisis,t-1} + \beta_{i2} \times DD_{fin-crisis,t-1} \\ + \sum_{j=1}^{J} \beta_{ij} \times control_{j,t-1} + Industry_{dummy} + Time_{dummy} + \varepsilon_{i,t} \quad (16)$$

where $DD_{i,t}$ represents the distance-to-default for industry $i$, $DD_{fin,t-1}$ is the distance-to-default for the financial sector lagged one period, $control_{j,t-1}$ are a set of control variables lagged one period,[28] and $\varepsilon_{i,t}$ is the residual.

We include industry and time dummies to avoid estimation bias, and we estimate the coefficients by means of a Prais-Winsten regression robust to heteroskedasticity and contemporaneous correlation across panels. If there is significant impact of the lagged financial sector's DD on industry $i$'s DD we should expect that $\beta_{i1} > 0$ in (15). Also, if the financial crisis increases the impact of these spillovers we should expect that $\beta_{i2} > \beta_{i1}$ in (16).

---

[26] The Basel Committee on Banking Supervision (1999) considers Merton's DD model as an industry standard risk. We calculate DD in a usual way as many literatures did, which is based on Merton's model (1974). The detail of the calculation is available on request.

[27] The variable $DD_{fin-non-crisis}$ is equal to $DD_{fin}$ multiplied by a dummy variable which is equal to one before the second quarter of 2007 and after the first quarter of 2009 and equals zero otherwise while $DD_{fin-crisis}$ variable is obtained by multiplying the $DD_{fin}$ by a dummy variable which equals one after the second quarter of 2007 and before the first quarter of 2009 and equals zero otherwise.

[28] Industries' control variables are: the volatility of profitability (VOLP), leverage, earnings per share (EP), and SIZE (defined as market value instead of total assets to avoid any collinearity problem) for each industry.

The empirical results are reported in Table 10. First, there is a significant and positive relation between the industries' DD and financial sector's DD. The control variables are significant and with the expected sign.[29] Comparing with results of Model 1, the Model 2 of Table 10 allows us to investigate whether the influence of tail risk spillover is stronger during the crisis period. The evidence shows a larger estimated coefficient on crisis-specific DD, revealing that tail risk spillover is stronger during crisis periods. Overall, results are supportive of our hypothesis 2a.

[INSERT TABLE 10 HERE]

*6.2. The impact of extreme returns*

Given that we find extreme negative returns in some health-care related industries (Hospitals, Nursing, etc.) that might distort the analysis, we re-examine the relationship between CCX and industry characteristics by excluding these sectors. We only have to re-test the cases based on all industries and concentrated industries since these suspicious sectors are all identified as concentrated ones in our study. Generally speaking, our main analysis remains the same.[30]

*6.3. Reverse causality: Volatility*

In order to test for reverse causality in volatility spillovers from real economy sectors to the financial sector, we switch *i*, and FIN in equation (1)-(5), and re-estimate this set of equations for each industry *i*. The results hardly provide supportive evidence of significant volatility spillovers from non-financial sectors to the financial industry in the full sample period. However in the crisis period there are many cases of significant volatility spillovers from real economy sectors to the financial sector. As an additional test, we also use the 30 industry daily data from Fama-French website. The results are similar. Therefore it seems that in crisis periods, there is a complex feedback mechanism of volatility spillovers between the financial sector and many real economy sectors. However it should be remarked that the initial shock to the economic system was originated in the financial sector.

---

[29] The signs of these coefficients are consistent with simple economic intuition; for example, smaller firms tend to use more short-term debt than larger firms, which make the whole industry riskier and more prone to financial distress. An industry with higher volatility of profitability and higher leverage bears more risk and thereby decreases the DD (higher default probability). Based on the above statement, higher volatility of profitability and leverage will result in lower DD (higher default probability), whereas higher earnings per share and size will cause higher DD (lower default probability).

[30] To save space, the results in Section 6.2, 6.3, and 6.4 are not included here but are available upon request.

*6.4.Reverse causality: Tail Risk*

Possible feedback effects of tail events between the financial and non-financial sectors are difficult to take into account in our baseline setting, because by its very nature the $CCX_t^i$ measure focuses on concurrent events. To shed light on possible feedback effects, we employ two additional CCX-based measures:

(i) $CCX_{(i,t,FIN,t-1)} = I_t^i(c) \times I_{t-1}^{FIN}(c), t = 1, \dots, T$ , which equals one if industry *i* has an exceedance on day *t* and the financial industry has an exceedance on day *t-1*.

(ii) $CCX_{(i,t-1,FIN,t)} = I_{t-1}^i(c) \times I_t^{FIN}(c), t = 1, \dots, T$ , which equals one if industry *i* has an exceedance on day "*t-1*" and the financial industry has an exceedance on day "*t*".

We hypothesize that, if there is a feedback effect and each sector (financial and non-financial) is causing each other, measures $CCX_{(i,t,FIN,t-1)}$ and $CCX_{(i,t-1,FIN,t)}$ should have similar likelihood. On the other hand, if tail risk events tend to manifest in the first place in the financial (non-financial) sector and they subsequently appear in the non-financial (financial) sector, measure $CCX_{(i,t,fin,t-1)}$ should have higher (lower) likelihood than measure $CCX_{(i,t-1,fin,t)}$. In order to test this idea, we compute likelihoods for both measures across the full sample period, the crisis period, and the non-crisis period. Then, we compute average values of probabilities of CCX on all industries across different time periods. The results (available on request) suggest that the likelihood of measure $CCX_{(i,t,fin,t-1)}$ is significantly higher than the likelihood of $CCX_{(i,t-1,fin,t)}$ during the full sample period and in the crisis period. Therefore, the evidence suggests that in crisis periods (and in the full sample) tail risk events appear in the first place in the financial sector and afterwards in the non-financial sectors. On the other hand, in the non-crisis period a feedback effect cannot be ruled out. Therefore our assumptions that, in crisis periods, extreme volatility episodes start within the financial sector and then spill over to other non-financial sectors seem to be borne by the data.

*6.5. Dynamic CCX*

To assess the possible dynamic nature of CCX, we re-computed this variable using a moving three-day window. That is, the CCX is defined now including financial sector's extreme returns within a moving three day window. Then we compute the likelihoods and re-estimate the base model regression of Eq. (10a) using this alternative measure as the dependent variable. We find that results are broadly consistent with our hypothesis 2a, 3a, 3b, and 3c.

## 7. Conclusion

In this paper, we propose a new approach to measure tail risk spillovers from one economic sector to other sectors. This new measure is the Conditional coexceedance or CCX. We apply our approach to the case of the financial sector and real economy industries in the US in the period from 2001 to 2011. We find large volatility and tail risk spillovers from the financial sector to many real economy sectors during this period. These spillovers are even stronger during the 2007-2009 financial crisis. Also we find some evidence suggesting that the higher the degree of competition the stronger these tail risk spillovers.

Three industry characteristics help to explain the size of tail spillovers. The net debt financing has a positive effect on the size whereas valuation and investment have a negative impact. The variable having the most relevant impact from the economic point of view is the valuation. Our results have implications for practitioners because they support the view that difficulties in the financial sector are, sooner or later, followed by large increases in uncertainty in many (but not all) industries and services. Furthermore our results add to the growing body of literature supporting the financial sector risk's role as a leading indicator of real economy overall risk and specially tail risk. Our results may also help financial regulators trying to evaluate the true overall economic cost of financial crises.

Looking forward, while recent research shows that firm' capital structure decisions depend on industry structure (for instance, Bradley et al., 1984; MacKay and Phillips, 2005), the extent to which product market structure may affect tail risk remains an open question. Given that we provide evidence linking product market characteristics to CCX this is an interesting avenue for future research. Another interesting future line of research is to explore the implications of our findings on loss aversion-based portfolio choice.

**Acknowledgments**

We thank Xin Zhao and other participants in the 3rd Financial Engineering and Banking Society (FEBS) 2013 Conference on Financial Regulation and Systemic Risk, for useful comments. Two anonymous referees provided many valuable suggestions which considerably improved the paper. The usual disclaimer applies. Peña acknowledges financial support from MCI grants ECO2009-12551 and ECO2012-35023.

# Appendix A

| NAICS code | Industry name | NAICS code | Industry name |
|---|---|---|---|
| 111 | Crop Production | 444 | Building Material and Garden Equipment and Supplies Dealers |
| 112 | Animal Production and Aquaculture | 445 | Food and Beverage Stores |
| 113 | Forestry and Logging | 446 | Health and Personal Care Stores |
| 211 | Oil and Gas Extraction | 447 | Gasoline Stations |
| 212 | Mining (except Oil and Gas) | 448 | Clothing and Clothing Accessories Stores |
| 213 | Support Activities for Mining | 451 | Sporting Goods, Hobby, Musical Instrument, and Book Stores |
| 221 | Utilities | 452 | General Merchandise Stores |
| 236 | Construction of Buildings | 453 | Miscellaneous Store Retailers |
| 237 | Heavy and Civil Engineering Construction | 454 | Nonstore Retailers |
| 238 | Specialty Trade Contractors | 481 | Air Transportation |
| 311 | Food Manufacturing | 482 | Rail Transportation |
| 312 | Beverage and Tobacco Product Manufacturing | 483 | Water Transportation |
| 313 | Textile Mills | 484 | Truck Transportation |
| 314 | Textile Product Mills | 486 | Pipeline Transportation |
| 315 | Apparel Manufacturing | 488 | Support Activities for Transportation |
| 316 | Leather and Allied Product Manufacturing | 492 | Couriers and Messengers |
| 321 | Wood Product Manufacturing | 493 | Warehousing and Storage |
| 322 | Paper Manufacturing | 511 | Publishing Industries (except Internet) |
| 323 | Printing and Related Support Activities | 512 | Motion Picture and Sound Recording Industries |
| 324 | Petroleum and Coal Products Manufacturing | 515 | Broadcasting (except Internet) |
| 325 | Chemical Manufacturing | 517 | Telecommunications |
| 326 | Plastics and Rubber Products Manufacturing | 518 | Data Processing, Hosting, and Related Services |
| 327 | Nonmetallic Mineral Product Manufacturing | 519 | Other Information Services |
| 331 | Primary Metal Manufacturing | 541 | Professional, Scientific, and Technical Services |
| 332 | Fabricated Metal Product Manufacturing | 561 | Administrative and Support Services |
| 333 | Machinery Manufacturing | 562 | Waste Management and Remediation Services |
| 334 | Computer and Electronic Product Manufacturing | 611 | Educational Services |
| 335 | Electrical Equipment, Appliance, and Component Manufacturing | 621 | Ambulatory Health Care Services |
| 336 | Transportation Equipment Manufacturing | 622 | Hospitals |
| 337 | Furniture and Related Product Manufacturing | 623 | Nursing and Residential Care Facilities |
| 339 | Miscellaneous Manufacturing | 711 | Performing Arts, Spectator Sports, and Related Industries |
| 423 | Merchant Wholesalers, Durable Goods | 713 | Amusement, Gambling, and Recreation Industries |
| 424 | Merchant Wholesalers, Nondurable Goods | 721 | Accommodation |
| 425 | Wholesale Electronic Markets and Agents and Brokers | 722 | Food Services and Drinking Places |
| 441 | Motor Vehicle and Parts Dealers | 811 | Repair and Maintenance |
| 442 | Furniture and Home Furnishings Stores | 812 | Personal and Laundry Services |
| 443 | Electronics and Appliance Stores | | |

**Appendix B**

**Classifying industries as competitive or concentrated**

Following Hoberg and Phillips (2010), we classify industries as competitive or concentrated using an indicator which combines Compustat data with Herfindahl index data from the Census Bureau (US Department of Commerce) and employee data from the Bureau of Labor statistics (BLS).

The fitted-HHI is computed by means of a two-step procedure. In the first step, for the manufacturing industries where we have information on their HHIs including both public and private firms for every five years, we regress the industry HHI (obtained from the Commerce Department) on three variables: the Compustat public-firm-only Herfindahl index, the average number of employees per firm using the BLS data (based on public and private firms), and the number of employees per firm for public firms using Compustat data. In the second step, we use the estimated coefficients from this regression to compute fitted HHI for all industries. This fitted HHI has the advantage of capturing the influence of both public and private firms. In this paper, we do not use these fitted HHI directly as an explanatory variable into any regression because of possible measurement errors. However, we regard the highest 25% (10%) of fitted HHI to correspond to concentrated industries, and those with lowest 25% (10%) to correspond to competitive industries.

**Table 1**
Descriptive statistics of industry-level returns over the full sample period.

| Industry Name | Mean | Median | Maximum | Minimum | Std. Dev. | Skewness | Kurtosis | Jarque-Bera | p-Value | # of firms[a] |
|---|---|---|---|---|---|---|---|---|---|---|
| Crop Production | 0.001 | 0.001 | 0.188 | -0.155 | 0.021 | 0.246 | 11.286 | 7942.86 | 0 | 9.96 |
| Animal Production and Aquaculture | 0 | 0.001 | 0.129 | -0.149 | 0.021 | -0.452 | 11.014 | 7497.95 | 0 | 2.59 |
| Forestry and Logging | 0.001 | 0 | 0.241 | -0.146 | 0.021 | 0.96 | 17.413 | 24375.91 | 0 | 2.60 |
| Oil and Gas Extraction | 0.001 | 0.002 | 0.206 | -0.166 | 0.022 | -0.174 | 12.055 | 9466.69 | 0 | 138.83 |
| Mining (except Oil and Gas) | 0.001 | 0.002 | 0.209 | -0.135 | 0.023 | 0.117 | 9.867 | 5442.66 | 0 | 95.45 |
| Support Activities for Mining | 0.001 | 0.001 | 0.23 | -0.176 | 0.025 | -0.209 | 9.041 | 4227.18 | 0 | 35.93 |
| Utilities | 0.001 | 0.001 | 0.147 | -0.086 | 0.013 | 0.454 | 16.244 | 20316.47 | 0 | 131.36 |
| Construction of Buildings | 0.001 | 0 | 0.16 | -0.126 | 0.026 | 0.196 | 5.595 | 794.36 | 0 | 25.41 |
| Heavy and Civil Engineering Construction | 0.001 | 0.001 | 0.139 | -0.176 | 0.023 | -0.242 | 7.916 | 2813.71 | 0 | 25.02 |
| Specialty Trade Contractors | 0.001 | 0.002 | 0.19 | -0.142 | 0.024 | -0.046 | 10.163 | 5916.09 | 0 | 11.99 |
| Food Manufacturing | 0 | 0.001 | 0.093 | -0.069 | 0.01 | 0.141 | 11.607 | 8549.41 | 0 | 74.57 |
| Beverage and Tobacco Product Manufacturing | 0.001 | 0.001 | 0.112 | -0.084 | 0.011 | -0.14 | 10.695 | 6835.27 | 0 | 40.76 |
| Textile Mills | 0.001 | 0.001 | 0.178 | -0.13 | 0.021 | 0.174 | 8.811 | 3907.22 | 0 | 8.48 |
| Textile Product Mills | 0.001 | 0 | 0.285 | -0.126 | 0.024 | 0.838 | 13.183 | 12277.96 | 0 | 4.09 |
| Apparel Manufacturing | 0.001 | 0.001 | 0.122 | -0.093 | 0.017 | 0.078 | 7.028 | 1873.62 | 0 | 42.63 |
| Leather and Allied Product Manufacturing | 0.001 | 0.001 | 0.13 | -0.118 | 0.017 | 0.134 | 8.439 | 3418.76 | 0 | 20.45 |
| Wood Product Manufacturing | 0.001 | 0.001 | 0.126 | -0.154 | 0.021 | -0.082 | 7.609 | 2451.83 | 0 | 18.53 |
| Paper Manufacturing | 0 | 0.001 | 0.087 | -0.097 | 0.014 | -0.119 | 7.835 | 2701.93 | 0 | 39.85 |
| Printing and Related Support Activities | 0.001 | 0.001 | 0.142 | -0.113 | 0.019 | 0.156 | 10.64 | 6741.57 | 0 | 19.40 |
| Petroleum and Coal Products Manufacturing | 0.001 | 0.001 | 0.177 | -0.136 | 0.017 | 0.168 | 14.5 | 15259.19 | 0 | 36.56 |
| Chemical Manufacturing | 0 | 0.001 | 0.109 | -0.067 | 0.012 | 0.055 | 9.63 | 5069.22 | 0 | 474.89 |
| Plastics and Rubber Products Manufacturing | 0.001 | 0.001 | 0.101 | -0.109 | 0.017 | -0.058 | 6.889 | 1745.08 | 0 | 36.63 |
| Nonmetallic Mineral Product Manufacturing | 0.001 | 0.001 | 0.145 | -0.119 | 0.022 | 0.172 | 8.508 | 3510.92 | 0 | 21.19 |
| Primary Metal Manufacturing | 0.001 | 0.001 | 0.247 | -0.179 | 0.026 | 0.055 | 11.86 | 9051.77 | 0 | 56.57 |
| Fabricated Metal Product Manufacturing | 0.001 | 0.001 | 0.104 | -0.097 | 0.016 | -0.019 | 7.582 | 2420.51 | 0 | 58.73 |
| Machinery Manufacturing | 0.001 | 0.001 | 0.138 | -0.117 | 0.019 | 0.03 | 7.525 | 2361.32 | 0 | 169.01 |
| Computer and Electronic Product Manufacturing | 0.001 | 0.001 | 0.171 | -0.086 | 0.02 | 0.502 | 8.351 | 3416.61 | 0 | 627.70 |
| Electrical Equipment, Appliance, and Component Manufacturing | 0.001 | 0.001 | 0.138 | -0.124 | 0.018 | 0.039 | 8.485 | 3468.93 | 0 | 82.71 |
| Transportation Equipment Manufacturing | 0.001 | 0.001 | 0.125 | -0.115 | 0.016 | -0.117 | 8.858 | 3963.19 | 0 | 107.14 |
| Furniture and Related Product Manufacturing | 0.001 | 0.001 | 0.12 | -0.112 | 0.019 | 0.132 | 7.91 | 2787.63 | 0 | 24.58 |
| Miscellaneous Manufacturing | 0.001 | 0.001 | 0.119 | -0.072 | 0.012 | -0.054 | 10.297 | 6139.63 | 0 | 135.39 |
| Merchant Wholesalers, Durable Goods | 0.001 | 0.001 | 0.086 | -0.088 | 0.015 | -0.021 | 6.516 | 1425.62 | 0 | 88.63 |
| Merchant Wholesalers, Nondurable Goods | 0.001 | 0.001 | 0.123 | -0.085 | 0.013 | -0.046 | 10.934 | 7258.44 | 0 | 53.74 |
| Wholesale Electronic Markets and Agents and Brokers | 0.001 | 0 | 0.176 | -0.268 | 0.023 | -0.459 | 20.265 | 34461.76 | 0 | 2.26 |
| Motor Vehicle and Parts Dealers | 0.001 | 0.001 | 0.116 | -0.097 | 0.017 | 0.458 | 7.968 | 2941.8 | 0 | 19.26 |
| Furniture and Home Furnishings Stores | 0.001 | 0 | 0.209 | -0.115 | 0.022 | 0.753 | 9.057 | 4491.15 | 0 | 7.72 |
| Electronics and Appliance Stores | 0.001 | 0.001 | 0.167 | -0.171 | 0.023 | 0.171 | 8.406 | 3382.49 | 0 | 10.18 |
| Building Material and Garden Equipment and Supplies Dealers | 0 | 0 | 0.15 | -0.108 | 0.02 | 0.476 | 7.902 | 2874.39 | 0 | 5.03 |

| Industry | | | | | | | | | | |
|---|---|---|---|---|---|---|---|---|---|---|
| Food and Beverage Stores | 0 | 0 | 0.08 | -0.089 | 0.015 | -0.354 | 6.518 | 1484.35 | 0 | 21.20 |
| Health and Personal Care Stores | 0 | 0 | 0.133 | -0.092 | 0.015 | -0.05 | 9.232 | 4478.15 | 0 | 14.15 |
| Gasoline Stations | 0.001 | 0.001 | 0.188 | -0.163 | 0.021 | 0.348 | 9.669 | 5182.87 | 0 | 4.33 |
| Clothing and Clothing Accessories Stores | 0.001 | 0.001 | 0.133 | -0.121 | 0.019 | 0.183 | 6.95 | 1814.42 | 0 | 54.32 |
| Sporting Goods, Hobby, Musical Instrument, and Book Stores | 0.001 | 0.001 | 0.124 | -0.101 | 0.02 | 0.322 | 6.593 | 1536.61 | 0 | 17.31 |
| General Merchandise Stores | 0 | 0 | 0.118 | -0.083 | 0.014 | 0.371 | 7.766 | 2682.41 | 0 | 25.15 |
| Miscellaneous Store Retailers | 0.001 | 0 | 0.166 | -0.12 | 0.021 | 0.537 | 8.688 | 3862.57 | 0 | 13.11 |
| Nonstore Retailers | 0.002 | 0.001 | 0.195 | -0.114 | 0.023 | 0.766 | 10.045 | 5992.94 | 0 | 33.25 |
| Air Transportation | 0.001 | 0 | 0.162 | -0.311 | 0.025 | -0.322 | 13.839 | 13591.9 | 0 | 28.28 |
| Rail Transportation | 0.001 | 0.001 | 0.111 | -0.107 | 0.018 | -0.043 | 6.381 | 1318.74 | 0 | 12.97 |
| Water Transportation | 0.001 | 0.001 | 0.15 | -0.25 | 0.021 | -0.638 | 15.011 | 16819.24 | 0 | 37.40 |
| Truck Transportation | 0.001 | 0 | 0.105 | -0.116 | 0.02 | 0.108 | 5.561 | 761.31 | 0 | 24.79 |
| Pipeline Transportation | 0.001 | 0.001 | 0.212 | -0.1 | 0.015 | 0.374 | 22.74 | 44990.19 | 0 | 21.81 |
| Support Activities for Transportation | 0.001 | 0.001 | 0.102 | -0.101 | 0.018 | 0.04 | 6.229 | 1202.96 | 0 | 13.80 |
| Couriers and Messengers | 0 | 0 | 0.079 | -0.097 | 0.016 | 0.07 | 7.083 | 1924.53 | 0 | 8.35 |
| Warehousing and Storage | 0.001 | 0.001 | 0.131 | -0.108 | 0.014 | 0.26 | 14.762 | 15982.18 | 0 | 3.08 |
| Publishing Industries (except Internet) | 0.001 | 0 | 0.152 | -0.085 | 0.018 | 0.535 | 9.678 | 5273.83 | 0 | 273.65 |
| Motion Picture and Sound Recording Industries | 0.001 | 0 | 0.108 | -0.092 | 0.017 | 0.214 | 7.172 | 2028.13 | 0 | 18.94 |
| Broadcasting (except Internet) | 0 | 0 | 0.154 | -0.12 | 0.019 | 0.271 | 10.356 | 6271.68 | 0 | 63.26 |
| Telecommunications | 0.001 | 0.001 | 0.142 | -0.086 | 0.015 | 0.443 | 11.054 | 7570.13 | 0 | 156.64 |
| Data Processing, Hosting, and Related Services | 0.001 | 0.001 | 0.114 | -0.098 | 0.014 | 0.139 | 8.429 | 3406.36 | 0 | 44.25 |
| Other Information Services | 0.001 | 0.001 | 0.162 | -0.118 | 0.022 | 0.373 | 6.943 | 1856.62 | 0 | 53.81 |
| Professional, Scientific, and Technical Services | 0.001 | 0.001 | 0.103 | -0.072 | 0.015 | 0.277 | 7.802 | 2693.74 | 0 | 248.67 |
| Administrative and Support Services | 0.001 | 0.001 | 0.095 | -0.094 | 0.016 | 0.04 | 7.042 | 1884.53 | 0 | 85.91 |
| Waste Management and Remediation Services | 0.001 | 0.001 | 0.151 | -0.096 | 0.015 | 0.149 | 10.265 | 6094.81 | 0 | 19.46 |
| Educational Services | 0.001 | 0.001 | 0.104 | -0.13 | 0.018 | -0.167 | 7.331 | 2175.22 | 0 | 22.45 |
| Ambulatory Health Care Services | 0.001 | 0.002 | 0.125 | -0.148 | 0.017 | -0.755 | 11.155 | 7929.94 | 0 | 32.16 |
| Hospitals | 0.001 | 0.001 | 0.231 | -0.251 | 0.028 | 0.064 | 13.016 | 11569.13 | 0 | 5.48 |
| Nursing and Residential Care Facilities | 0.001 | 0.001 | 0.42 | -0.235 | 0.028 | 1.906 | 49.086 | 246548.2 | 0 | 7.36 |
| Performing Arts, Spectator Sports, and Related Industries | 0 | 0 | 0.132 | -0.113 | 0.018 | -0.017 | 9.035 | 4198.78 | 0 | 7.43 |
| Amusement, Gambling, and Recreation Industries | 0.001 | 0.001 | 0.146 | -0.132 | 0.02 | 0.237 | 9.939 | 5577.67 | 0 | 21.81 |
| Accommodation | 0.001 | 0.001 | 0.234 | -0.199 | 0.026 | 0.322 | 11.849 | 9074.63 | 0 | 29.55 |
| Food Services and Drinking Places | 0.001 | 0.001 | 0.093 | -0.082 | 0.013 | 0.049 | 6.182 | 1168.22 | 0 | 67.25 |
| Repair and Maintenance | 0.001 | 0 | 0.096 | -0.117 | 0.021 | 0.148 | 5.06 | 499.36 | 0 | 2.91 |
| Personal and Laundry Services | 0.001 | 0.001 | 0.168 | -0.104 | 0.016 | 0.356 | 10.866 | 7192.03 | 0 | 14.85 |
| Finance | 0.001 | 0.001 | 0.149 | -0.13 | 0.019 | 0.482 | 13.813 | 13588.04 | 0 | 1796.00 |

*Notes:* Descriptive statistics of the daily returns for each industry during the full sample period of 2001–2011, totaling 2767 daily observations for each industry.

[a] The average number of firms that are contained in each industry.

**Table 2**
Descriptive statistics of industry-level returns during crisis period.

| Industry Name | Mean | Median | Maximum | Minimum | Std. Dev. | Skewness | Kurtosis | Jarque-Bera | p-Value | # of firms |
|---|---|---|---|---|---|---|---|---|---|---|
| Crop Production | 0.001 | 0.002 | 0.188 | -0.155 | 0.037 | 0.293 | 6.821 | 274.612 | 0 | 7.58 |
| Animal Production and Aquaculture | 0 | 0 | 0.093 | -0.149 | 0.023 | -1.152 | 12.127 | 1628.216 | 0 | 2.71 |
| Forestry and Logging | 0 | 0 | 0.241 | -0.146 | 0.040 | 0.988 | 8.414 | 610.327 | 0 | 2.00 |
| Oil and Gas Extraction | 0 | 0.003 | 0.206 | -0.166 | 0.037 | -0.023 | 7.880 | 437.618 | 0 | 157.99 |
| Mining (except Oil and Gas) | 0.001 | 0.002 | 0.209 | -0.135 | 0.038 | 0.274 | 6.496 | 230.041 | 0 | 111.47 |
| Support Activities for Mining | -0.001 | 0.001 | 0.230 | -0.176 | 0.038 | -0.153 | 8.547 | 567.016 | 0 | 37.89 |
| Utilities | 0 | 0.001 | 0.147 | -0.086 | 0.021 | 0.939 | 11.791 | 1484.998 | 0 | 126.24 |
| Construction of Buildings | 0 | -0.001 | 0.160 | -0.126 | 0.040 | 0.264 | 3.955 | 21.888 | 0 | 25.83 |
| Heavy and Civil Engineering Construction | -0.001 | 0 | 0.139 | -0.176 | 0.036 | -0.296 | 6.278 | 203.842 | 0 | 26.76 |
| Specialty Trade Contractors | 0 | 0 | 0.190 | -0.142 | 0.039 | 0.146 | 6.581 | 237.244 | 0 | 13.74 |
| Food Manufacturing | 0 | 0 | 0.093 | -0.069 | 0.016 | 0.532 | 8.970 | 675.805 | 0 | 70.46 |
| Beverage and Tobacco Product Manufacturing | 0 | 0 | 0.112 | -0.078 | 0.015 | 0.403 | 12.55 | 1687.734 | 0 | 38.24 |
| Textile Mills | -0.002 | -0.001 | 0.105 | -0.130 | 0.027 | -0.140 | 5.739 | 139.329 | 0 | 6.07 |
| Textile Product Mills | -0.002 | -0.004 | 0.163 | -0.126 | 0.034 | 0.434 | 5.655 | 143.380 | 0 | 4.00 |
| Apparel Manufacturing | -0.001 | -0.002 | 0.122 | -0.093 | 0.027 | 0.188 | 4.729 | 57.546 | 0 | 38.68 |
| Leather and Allied Product Manufacturing | -0.001 | -0.001 | 0.130 | -0.118 | 0.028 | 0.340 | 5.340 | 109.078 | 0 | 17.08 |
| Wood Product Manufacturing | -0.001 | 0.001 | 0.126 | -0.154 | 0.033 | -0.074 | 5.289 | 96.669 | 0 | 17.01 |
| Paper Manufacturing | -0.001 | 0 | 0.087 | -0.097 | 0.020 | -0.029 | 6.373 | 209.157 | 0 | 34.42 |
| Printing and Related Support Activities | -0.002 | -0.001 | 0.142 | -0.113 | 0.031 | 0.384 | 6.015 | 177.797 | 0 | 14.49 |
| Petroleum and Coal Products Manufacturing | 0 | 0.002 | 0.177 | -0.136 | 0.028 | 0.397 | 9.774 | 854.685 | 0 | 34.98 |
| Chemical Manufacturing | 0 | 0.001 | 0.109 | -0.067 | 0.017 | 0.319 | 9.306 | 738.218 | 0 | 494.76 |
| Plastics and Rubber Products Manufacturing | -0.001 | -0.002 | 0.086 | -0.109 | 0.027 | -0.048 | 4.361 | 34.198 | 0 | 26.22 |
| Nonmetallic Mineral Product Manufacturing | -0.001 | -0.001 | 0.145 | -0.119 | 0.034 | 0.327 | 5.667 | 138.526 | 0 | 19.01 |
| Primary Metal Manufacturing | -0.001 | 0 | 0.247 | -0.179 | 0.045 | 0.201 | 7.032 | 301.744 | 0 | 48.76 |
| Fabricated Metal Product Manufacturing | -0.001 | 0 | 0.104 | -0.097 | 0.025 | 0.007 | 4.853 | 63.108 | 0 | 51.85 |
| Machinery Manufacturing | -0.001 | 0.001 | 0.138 | -0.117 | 0.029 | -0.030 | 6.178 | 185.699 | 0 | 149.11 |
| Computer and Electronic Product Manufacturing | 0 | 0.001 | 0.118 | -0.086 | 0.023 | 0.268 | 5.860 | 155.607 | 0 | 598.00 |
| Electrical Equipment, Appliance, and Component Manufacturing | -0.001 | 0 | 0.138 | -0.124 | 0.028 | 0.107 | 6.403 | 213.63 | 0 | 79.67 |
| Transportation Equipment Manufacturing | -0.001 | 0 | 0.125 | -0.093 | 0.025 | 0.227 | 6.469 | 224.951 | 0 | 101.92 |
| Furniture and Related Product Manufacturing | -0.001 | -0.001 | 0.120 | -0.112 | 0.031 | 0.290 | 4.759 | 63.039 | 0 | 20.45 |
| Miscellaneous Manufacturing | 0 | 0 | 0.119 | -0.072 | 0.018 | 0.136 | 9.425 | 759.924 | 0 | 125.51 |
| Merchant Wholesalers, Durable Goods | 0 | 0.001 | 0.086 | -0.088 | 0.023 | -0.110 | 4.885 | 66.179 | 0 | 79.39 |
| Merchant Wholesalers, Nondurable Goods | -0.001 | 0 | 0.123 | -0.078 | 0.020 | 0.300 | 9.148 | 701.022 | 0 | 51.40 |
| Wholesale Electronic Markets and Agents and Brokers | 0 | -0.001 | 0.128 | -0.106 | 0.024 | 0.151 | 6.666 | 248.579 | 0 | 2.00 |
| Motor Vehicle and Parts Dealers | 0 | -0.002 | 0.116 | -0.084 | 0.027 | 0.514 | 4.863 | 83.194 | 0 | 17.66 |
| Furniture and Home Furnishings Stores | 0 | -0.002 | 0.140 | -0.103 | 0.031 | 0.620 | 4.905 | 94.893 | 0 | 7.55 |
| Electronics and Appliance Stores | 0 | 0 | 0.149 | -0.098 | 0.030 | 0.541 | 6.040 | 191.355 | 0 | 8.14 |
| Building Material and Garden Equipment and Supplies Dealers | -0.001 | -0.003 | 0.150 | -0.084 | 0.029 | 0.717 | 5.209 | 127.394 | 0 | 4.47 |

| Industry | | | | | | | | | | |
|---|---|---|---|---|---|---|---|---|---|---|
| Food and Beverage Stores | -0.001 | -0.001 | 0.080 | -0.085 | 0.021 | -0.021 | 4.614 | 47.902 | 0 | 15.05 |
| Health and Personal Care Stores | 0 | -0.001 | 0.133 | -0.092 | 0.021 | 0.424 | 9.217 | 723.419 | 0 | 13.08 |
| Gasoline Stations | 0 | 0.001 | 0.188 | -0.140 | 0.029 | 0.675 | 9.686 | 854.968 | 0 | 5.42 |
| Clothing and Clothing Accessories Stores | 0 | -0.002 | 0.117 | -0.099 | 0.029 | 0.228 | 4.300 | 34.881 | 0 | 53.05 |
| Sporting Goods, Hobby, Musical Instrument, and Book Stores | -0.001 | -0.003 | 0.112 | -0.101 | 0.032 | 0.386 | 4.199 | 37.356 | 0 | 16.16 |
| General Merchandise Stores | 0 | -0.001 | 0.117 | -0.083 | 0.021 | 0.425 | 6.230 | 204.948 | 0 | 20.03 |
| Miscellaneous Store Retailers | -0.001 | -0.003 | 0.139 | -0.120 | 0.031 | 0.567 | 5.680 | 155.645 | 0 | 12.26 |
| Nonstore Retailers | 0 | -0.002 | 0.126 | -0.114 | 0.031 | 0.478 | 5.562 | 137.404 | 0 | 31.30 |
| Air Transportation | 0 | -0.001 | 0.162 | -0.116 | 0.039 | 0.428 | 4.446 | 51.870 | 0 | 28.27 |
| Rail Transportation | 0 | 0.001 | 0.094 | -0.107 | 0.027 | -0.124 | 4.426 | 38.506 | 0 | 10.04 |
| Water Transportation | -0.001 | -0.001 | 0.150 | -0.113 | 0.032 | 0.104 | 5.441 | 110.309 | 0 | 46.86 |
| Truck Transportation | 0 | -0.002 | 0.105 | -0.116 | 0.029 | 0.118 | 4.270 | 30.656 | 0 | 21.45 |
| Pipeline Transportation | 0 | 0 | 0.212 | -0.100 | 0.023 | 1.313 | 21.13 | 6166.848 | 0 | 26.41 |
| Support Activities for Transportation | 0 | 0 | 0.102 | -0.101 | 0.027 | 0.097 | 4.752 | 57.076 | 0 | 15.20 |
| Couriers and Messengers | -0.001 | -0.002 | 0.079 | -0.097 | 0.024 | 0.108 | 5.122 | 83.641 | 0 | 7.54 |
| Warehousing and Storage | 0 | 0 | 0.131 | -0.108 | 0.024 | 0.224 | 9.135 | 695.178 | 0 | 2.83 |
| Publishing Industries (except Internet) | 0 | -0.001 | 0.152 | -0.085 | 0.024 | 0.627 | 8.071 | 501.349 | 0 | 211.78 |
| Motion Picture and Sound Recording Industries | -0.001 | -0.002 | 0.108 | -0.09 | 0.024 | 0.299 | 5.544 | 125.494 | 0 | 16.42 |
| Broadcasting (except Internet) | -0.001 | -0.001 | 0.154 | -0.111 | 0.027 | 0.584 | 8.941 | 673.678 | 0 | 62.17 |
| Telecommunications | 0 | 0 | 0.142 | -0.086 | 0.023 | 0.717 | 8.792 | 654.228 | 0 | 145.73 |
| Data Processing, Hosting, and Related Services | 0 | 0 | 0.114 | -0.086 | 0.020 | 0.232 | 7.451 | 368.014 | 0 | 30.63 |
| Other Information Services | 0 | -0.001 | 0.135 | -0.106 | 0.026 | 0.483 | 6.786 | 280.492 | 0 | 63.70 |
| Professional, Scientific, and Technical Services | 0 | 0 | 0.088 | -0.071 | 0.020 | 0.238 | 5.577 | 126.222 | 0 | 217.01 |
| Administrative and Support Services | -0.001 | -0.001 | 0.095 | -0.094 | 0.025 | 0.144 | 4.842 | 63.857 | 0 | 73.70 |
| Waste Management and Remediation Services | 0 | 0 | 0.151 | -0.074 | 0.021 | 0.780 | 10.269 | 1015.623 | 0 | 16.16 |
| Educational Services | 0.001 | 0 | 0.094 | -0.113 | 0.025 | 0.0790 | 4.872 | 64.834 | 0 | 20.33 |
| Ambulatory Health Care Services | -0.001 | 0 | 0.125 | -0.148 | 0.026 | -0.885 | 9.427 | 816.685 | 0 | 26.16 |
| Hospitals | 0 | -0.002 | 0.231 | -0.251 | 0.045 | -0.03 | 8.297 | 515.563 | 0 | 3.56 |
| Nursing and Residential Care Facilities | -0.003 | -0.003 | 0.420 | -0.235 | 0.053 | 1.815 | 22.447 | 7191.117 | 0 | 4.63 |
| Performing Arts, Spectator Sports, and Related Industries | -0.002 | -0.003 | 0.132 | -0.113 | 0.029 | 0.161 | 5.473 | 114.242 | 0 | 7.34 |
| Amusement, Gambling, and Recreation Industries | -0.002 | -0.002 | 0.146 | -0.111 | 0.032 | 0.565 | 6.439 | 240.770 | 0 | 21.85 |
| Accommodation | -0.002 | -0.001 | 0.234 | -0.136 | 0.043 | 0.574 | 6.555 | 256.529 | 0 | 22.66 |
| Food Services and Drinking Places | 0 | 0 | 0.093 | -0.082 | 0.020 | 0.169 | 4.799 | 61.579 | 0 | 62.69 |
| Repair and Maintenance | 0 | -0.002 | 0.092 | -0.117 | 0.031 | 0.219 | 3.458 | 7.386 | 0.025 | 2.00 |
| Personal and Laundry Services | -0.001 | -0.001 | 0.083 | -0.104 | 0.023 | -0.077 | 5.565 | 121.363 | 0 | 14.32 |
| Finance | -0.001 | -0.003 | 0.149 | -0.130 | 0.035 | 0.455 | 6.004 | 180.977 | 0 | 1796.48 |

*Notes:* Descriptive statistics of the daily returns for each industry in the crisis period from July 2007 until March 2009, totaling 441 daily observations for each industry.

[a] The average number of firms that are contained in each industry.

**Table 3**
Descriptive statistics of the dependent and explanatory variables.

| | CCX[a] | NET_D_I[b] | VAL_I[c] | INV_I[d] | VOLP[e] | LEV[f] | DEBT_COST[g] | EP[h] | NI[i] | SIZE[j] |
|---|---|---|---|---|---|---|---|---|---|---|
| *Panel A: from January 2001 to December 2011 (# of observations: 3212)* | | | | | | | | | | |
| Mean | 1.640 | 0.008 | 0.001 | 0.046 | 0.027 | 0.212 | 0.230 | -0.054 | -0.002 | 6.068 |
| Median | 0 | 0.004 | 0.003 | 0.011 | 0.008 | 0.198 | 0.214 | -0.006 | 0.004 | 5.945 |
| Maximum | 20 | 0.269 | 1.969 | 24.633 | 0.480 | 0.790 | 0.696 | 7.675 | 0.709 | 9.935 |
| Minimum | 0 | -0.121 | -2.114 | -6.221 | 0 | 0.023 | 0.023 | -6.238 | -0.357 | 2.726 |
| Std. Dev. | 3.192 | 0.028 | 0.253 | 0.952 | 0.046 | 0.095 | 0.116 | 0.279 | 0.031 | 0.984 |
| *Panel B: from July 2007 to March 2009 (# of observations: 511)* | | | | | | | | | | |
| Mean | 5.511 | 0.017 | -0.170 | 0.099 | 0.037 | 0.217 | 0.244 | -0.112 | -0.007 | 6.216 |
| Median | 3 | 0.012 | -0.167 | 0.010 | 0.012 | 0.207 | 0.229 | -0.019 | 0.001 | 6.019 |
| Maximum | 20 | 0.218 | 1.969 | 24.633 | 0.479 | 0.482 | 0.651 | 0.113 | 0.076 | 9.935 |
| Minimum | 0 | -0.075 | -1.285 | -5.122 | 0 | 0.038 | 0.024 | -6.238 | -0.193 | 4.067 |
| Std. Dev. | 5.461 | 0.031 | 0.301 | 1.599 | 0.060 | 0.098 | 0.121 | 0.399 | 0.029 | 0.919 |

*Notes:* Summary statistics for variables employed in Eq. (10a) and Eq. (10b).

[a] Conditional coexceedance is computed as detailed in Section 3.2.1.

[b] net debt financing is computed as detailed in Section 4.2.1.

[c] spread from a normative value is computed as detailed in Section 4.2.1.

[d] spread from a normative investment is computed as detailed in Section 4.2.1.

[e] volatility of profitability is computed as detailed in Section 4.2.1.

[f] leverage is computed as detailed in Section 4.2.1.

[g] industry debt cost: long term debt divided by the sum of long term debt and market equity.

[h] earnings per share.

[i] net income.

[j] logarithm of market capitalization.

**Table 4**
Volatility spillovers.

| Industry Name | $\gamma_1$ (coef.) | t-stat | $\gamma_2$ (coef.) | t-stat | $\gamma_1$ | $\gamma_2$ | Industry Name | $\gamma_1$ (coef.) | t-stat | $\gamma_2$ (coef.) | t-stat | $\gamma_1$ | $\gamma_2$ |
|---|---|---|---|---|---|---|---|---|---|---|---|---|---|
| Crop Production | 0.453 | 2.74 | 1.210 | 3.15 | c | c | Building Material and Garden Equipment and Supplies Dealers | 0.373 | 4.3 | 0.744 | 3.13 | c | c |
| Animal Production and Aquaculture | 0.172 | 1.15 | 0.060 | 0.77 | | | Food and Beverage Stores | 0.280 | 4.46 | 0.344 | 2.82 | c | c |
| Forestry and Logging | 0.441 | 3.18 | 1.182 | 2.85 | c | c | Health and Personal Care Stores | 0.313 | 4.69 | 0.217 | 2.32 | c | c |
| Oil and Gas Extraction | 0.363 | 3.68 | 0.380 | 1.67 | c | c | Gasoline Stations | 0.800 | 4.19 | 0.332 | 1.53 | c | |
| Mining (except Oil and Gas) | 0.292 | 1.53 | 0.930 | 2.55 | | c | Clothing and Clothing Accessories Stores | 0.238 | 1.56 | 0.751 | 2.74 | | c |
| Support Activities for Mining | 0.642 | 3.73 | 0.247 | 1.09 | c | | Sporting Goods, Hobby, Musical Instrument, and Book Stores | 0.256 | 3.74 | 0.543 | 2.58 | c | c |
| Utilities | 0.008 | 0.11 | 0.374 | 2.18 | | c | General Merchandise Stores | 0.149 | 2 | 0.403 | 2.76 | c | c |
| Construction of Buildings | 0.704 | 2.48 | 1.242 | 2.74 | c | c | Miscellaneous Store Retailers | 0.523 | 4.6 | 0.764 | 2.99 | c | c |
| Heavy and Civil Engineering Construction | 0.503 | 4.43 | 0.759 | 3 | c | c | Nonstore Retailers | 0.519 | 3.24 | 0.348 | 2.78 | c | c |
| Specialty Trade Contractors | 0.493 | 2.06 | 0.825 | 2.15 | c | c | Air Transportation | 2.350 | 7.33 | 2.388 | 2.85 | c | c |
| Food Manufacturing | 0.122 | 2.93 | 0.146 | 2.15 | c | c | Rail Transportation | 0.392 | 4.54 | 0.565 | 2.66 | c | c |
| Beverage and Tobacco Product Manufacturing | 0.352 | 5.99 | 0.006 | 0.1 | c | | Water Transportation | 0.564 | 4.47 | 0.657 | 2.02 | c | c |
| Textile Mills | 0.415 | 3.48 | 0.170 | 1.16 | c | | Truck Transportation | 0.702 | 4.97 | 1.043 | 3.46 | c | c |
| Textile Product Mills | 0.000 | 0 | 1.107 | 0.55 | | | Pipeline Transportation | 0.095 | 1.27 | 0.068 | 0.76 | | |
| Apparel Manufacturing | 0.305 | 4.19 | 0.616 | 2.73 | c | c | Support Activities for Transportation | 0.582 | 3.47 | 0.606 | 2.57 | c | c |
| Leather and Allied Product Manufacturing | 0.362 | 4.45 | 0.762 | 3.19 | c | c | Couriers and Messengers | 0.369 | 5.16 | 0.249 | 1.89 | c | c |
| Wood Product Manufacturing | 0.365 | 4.18 | 0.671 | 2.7 | c | c | Warehousing and Storage | 0.309 | 4.27 | 0.153 | 1.72 | c | c |
| Paper Manufacturing | 0.218 | 2.5 | 0.284 | 2.43 | c | c | Publishing Industries (except Internet) | 0.195 | 4.05 | 0.260 | 2.28 | c | c |
| Printing and Related Support Activities | 0.225 | 4.48 | 0.520 | 2.59 | c | c | Motion Picture and Sound Recording Industries | 0.262 | 4.26 | 0.226 | 1.95 | c | c |
| Petroleum and Coal Products Manufacturing | 0.271 | 2.36 | 0.545 | 2.4 | c | c | Broadcasting (except Internet) | 0.090 | 0.95 | 0.340 | 2.47 | | c |
| Chemical Manufacturing | 0.116 | 1.76 | 0.115 | 1.45 | c | | Telecommunications | 0.109 | 1.45 | 0.312 | 2.29 | | c |
| Plastics and Rubber Products Manufacturing | 0.408 | 4.97 | 0.779 | 3.05 | c | c | Data Processing, Hosting, and Related Services | 0.198 | 4.32 | 0.060 | 0.82 | c | |
| Nonmetallic Mineral Product Manufacturing | 0.455 | 2.53 | 1.023 | 2.95 | c | c | Other Information Services | 0.226 | 1.45 | 0.311 | 2.15 | | c |
| Primary Metal Manufacturing | 0.748 | 3.01 | 1.001 | 2.4 | c | c | Professional, Scientific, and Technical Services | 0.057 | 0.61 | 0.172 | 1.7 | | c |
| Fabricated Metal Product Manufacturing | 0.000 | 0 | 0.634 | 2.75 | | c | Administrative and Support Services | 0.237 | 4.25 | 0.461 | 2.72 | c | c |
| Machinery Manufacturing | 0.272 | 3.45 | 0.307 | 2.02 | | c | Waste Management and Remediation Services | 0.370 | 4.92 | 0.259 | 2.09 | c | c |
| Computer and Electronic Product Manufacturing | 0.177 | 3.24 | 0.236 | 2.04 | c | c | Educational Services | 0.225 | 3.34 | 0.296 | 2.87 | c | c |
| Electrical Equipment, Appliance, and Component Manufacturing | 0.324 | 4.39 | 0.380 | 2.23 | c | c | Ambulatory Health Care Services | 0.252 | 1.62 | 1.260 | 3.56 | | c |
| Transportation Equipment Manufacturing | 0.133 | 1.07 | 0.439 | 2.34 | | c | Hospitals | 0.413 | 4.68 | 1.267 | 3.81 | c | c |
| Furniture and Related Product Manufacturing | 0.337 | 4.43 | 0.852 | 3.05 | c | c | Nursing and Residential Care Facilities | 0.000 | 0 | 0.873 | 2.1 | | c |
| Miscellaneous Manufacturing | 0.182 | 2.55 | 0.169 | 1.77 | c | c | Performing Arts, Spectator Sports, and Related Industries | 0.606 | 6.42 | 0.936 | 3.29 | c | c |
| Merchant Wholesalers, Durable Goods | 0.276 | 4.35 | 0.257 | 2.18 | c | c | Amusement, Gambling, and Recreation Industries | 0.815 | 6.09 | 0.291 | 1.28 | c | |
| Merchant Wholesalers, Nondurable Goods | 0.150 | 1.88 | 0.172 | 1.73 | c | c | Accommodation | 0.792 | 5.85 | 0.715 | 1.65 | c | |
| Wholesale Electronic Markets and Agents and Brokers | 0.091 | 2.76 | 0.158 | 2.75 | c | c | Food Services and Drinking Places | 0.227 | 4.77 | 0.221 | 2.09 | c | c |
| Motor Vehicle and Parts Dealers | 0.248 | 2.11 | 0.661 | 2.87 | c | c | Repair and Maintenance | 0.899 | 4.75 | 1.081 | 3.65 | c | c |
| Furniture and Home Furnishings Stores | 0.000 | 0 | 1.131 | 3.17 | | c | Personal and Laundry Services | 0.172 | 1.41 | 0.180 | 1.32 | | |
| Electronics and Appliance Stores | 0.202 | 0.89 | 0.211 | 1.22 | | | | | | | | | |

*Notes:* The results of the volatility spillover by after implementing equations (1)-(5). We report estimated coefficients of $\gamma_1$ and $\gamma_2$ (scaled by 1000). The "*c*" stands for spillover during the whole sample and during the crisis period.

**Table 5**
The likelihoods of CCX.

| NAICS3 | Prob[a] | Prob$_{crisis}$[b] | Prob$_{non-crisis}$[c] | NAICS3 | Prob[a] | Prob$_{crisis}$[b] | Prob$_{non-crisis}$[c] |
|---|---|---|---|---|---|---|---|
| 111 | 0.0192 | 0.0726 | 0.0090 | 444 | 0.0235 | 0.0930 | 0.0103 |
| 112 | 0.0141 | 0.0385 | 0.0095 | 445 | 0.0231 | 0.0726 | 0.0138 |
| 113 | 0.0289 | 0.1134 | 0.0129 | 446 | 0.0210 | 0.0726 | 0.0112 |
| 211 | 0.0257 | 0.0794 | 0.0155 | 447 | 0.0231 | 0.0771 | 0.0129 |
| 212 | 0.0213 | 0.0726 | 0.0116 | 448 | 0.0260 | 0.0952 | 0.0129 |
| 213 | 0.0242 | 0.0816 | 0.0133 | 451 | 0.0289 | 0.0952 | 0.0163 |
| 221 | 0.0246 | 0.0884 | 0.0125 | 452 | 0.0199 | 0.0748 | 0.0095 |
| 236 | 0.0286 | 0.1111 | 0.0129 | 453 | 0.0264 | 0.0930 | 0.0138 |
| 237 | 0.0253 | 0.0884 | 0.0133 | 454 | 0.0195 | 0.0726 | 0.0095 |
| 238 | 0.0296 | 0.0998 | 0.0163 | 481 | 0.0249 | 0.0839 | 0.0138 |
| 311 | 0.0271 | 0.0862 | 0.0159 | 482 | 0.0278 | 0.0907 | 0.0159 |
| 312 | 0.0224 | 0.0748 | 0.0125 | 483 | 0.0325 | 0.1088 | 0.0181 |
| 313 | 0.0253 | 0.0771 | 0.0155 | 484 | 0.0235 | 0.0839 | 0.0120 |
| 314 | 0.0246 | 0.0794 | 0.0142 | 486 | 0.0206 | 0.0680 | 0.0116 |
| 315 | 0.0314 | 0.1066 | 0.0172 | 488 | 0.0260 | 0.0930 | 0.0133 |
| 316 | 0.0300 | 0.1066 | 0.0155 | 492 | 0.0275 | 0.0794 | 0.0176 |
| 321 | 0.0322 | 0.1043 | 0.0185 | 493 | 0.0224 | 0.0703 | 0.0133 |
| 322 | 0.0322 | 0.0975 | 0.0198 | 511 | 0.0202 | 0.0726 | 0.0103 |
| 323 | 0.0332 | 0.1202 | 0.0168 | 512 | 0.0300 | 0.0975 | 0.0172 |
| 324 | 0.0253 | 0.0930 | 0.0125 | 515 | 0.0275 | 0.0794 | 0.0176 |
| 325 | 0.0289 | 0.0862 | 0.0181 | 517 | 0.0271 | 0.0998 | 0.0133 |
| 326 | 0.0351 | 0.1088 | 0.0211 | 518 | 0.0278 | 0.0930 | 0.0155 |
| 327 | 0.0332 | 0.1066 | 0.0193 | 519 | 0.0170 | 0.0544 | 0.0099 |
| 331 | 0.0300 | 0.0952 | 0.0176 | 541 | 0.0275 | 0.0816 | 0.0172 |
| 332 | 0.0354 | 0.1202 | 0.0193 | 561 | 0.0354 | 0.1111 | 0.0211 |
| 333 | 0.0307 | 0.0907 | 0.0193 | 562 | 0.0249 | 0.0816 | 0.0142 |
| 334 | 0.0192 | 0.0635 | 0.0107 | 611 | 0.0188 | 0.0612 | 0.0107 |
| 335 | 0.0343 | 0.1066 | 0.0206 | 621 | 0.0206 | 0.0726 | 0.0107 |
| 336 | 0.0340 | 0.1020 | 0.0211 | 622 | 0.0173 | 0.0726 | 0.0069 |
| 337 | 0.0325 | 0.1111 | 0.0176 | 623 | 0.0242 | 0.0884 | 0.0120 |
| 339 | 0.0271 | 0.0816 | 0.0168 | 711 | 0.0282 | 0.1043 | 0.0138 |
| 423 | 0.0347 | 0.1066 | 0.0211 | 713 | 0.0304 | 0.1043 | 0.0163 |
| 424 | 0.0296 | 0.0884 | 0.0185 | 721 | 0.0293 | 0.1020 | 0.0155 |
| 425 | 0.0134 | 0.0408 | 0.0082 | 722 | 0.0253 | 0.0862 | 0.0138 |
| 441 | 0.0249 | 0.0884 | 0.0129 | 811 | 0.0181 | 0.0703 | 0.0082 |
| 442 | 0.0224 | 0.0794 | 0.0116 | 812 | 0.0282 | 0.0907 | 0.0163 |
| 443 | 0.0192 | 0.0703 | 0.0095 | | | | |
| Average value of all industries | | | | | | | |
| | 0.0261 | 0.0875 | 0.0144 | | | | |
| | | Difference[†] | 0.0730[†] | | | | |
| | | P-value[‡] | <0.0001[‡] | | | | |

*Notes:* The NAICS3 indicates the three digit code of NAICS.

[a] The likelihood of CCX over the whole sample period.

[b] The likelihood of CCX during the crisis period (from July 2007 to March 2009).

[c] The likelihood of CCX during the non-crisis period.

[†] The average difference between Prob$_{crisis}$, and Prob$_{non-crisis}$.

[‡] One-sided *p*-value of the Wilcoxon two-sample test.

**Table 6**

The likelihoods of CCX : Competitive and Concentrated Industries.

| Year | Prob[a] | Prob$_{cometitive}$[b] | Prob$_{concentrated}$[c] | Difference[†] | P-value[‡] |
|---|---|---|---|---|---|
| 2001 | 0.0163 | 0.0177 | 0.0141 | 0.0036 * | 0.0674 |
| 2002 | 0.0160 | 0.0168 | 0.0146 | 0.0022 | 0.1447 |
| 2003 | 0.0186 | 0.0192 | 0.0190 | 0.0002 | 0.4807 |
| 2004 | 0.0175 | 0.0201 | 0.0154 | 0.0047 ** | 0.0139 |
| 2005 | 0.0142 | 0.0141 | 0.0132 | 0.0009 | 0.2393 |
| 2006 | 0.0203 | 0.0210 | 0.0190 | 0.0020 | 0.2307 |
| 2007 | 0.0299 | 0.0297 | 0.0294 | 0.0003 | 0.3791 |
| 2008 | 0.0329 | 0.0358 | 0.0329 | 0.0029 * | 0.0821 |
| 2009 | 0.0269 | 0.0300 | 0.0256 | 0.0044 ** | 0.0156 |
| 2010 | 0.0328 | 0.0351 | 0.0328 | 0.0023 * | 0.0998 |
| 2011 | 0.0331 | 0.0340 | 0.0313 | 0.0027 * | 0.0971 |
| Total sample | 0.0235 | 0.0248 | 0.0225 | 0.0023 *** | 0.0092 |

*Notes:* We identify competitive and concentrated industries using the fitted HHI criteria when the fitted HHI is upper 25% and lower 25% respectively. For each year, the sample contains 73 industries, where 18 of them belong to competitive industries and 18 of them are concentrated industries.

[a] The average yearly likelihood of CCX for all industries.
[b] The average yearly likelihood of CCX for competitive industries.
[b] The average yearly likelihood of CCX for concentrated industries.
[†] The average difference between Prob$_{crisis}$, and Prob$_{non-crisis}$.
[‡] One-sided *p*-value of the Wilcoxon two-sample test.
* Significant level at 10%.
** Significant level at 5%.
*** Significant level at 1%

**Table 7**

CCX: The impact of industry characteristics.

| | Model 1 | | Model 2 | | Model 3 | | Model 4 | | Model 5 | |
|---|---|---|---|---|---|---|---|---|---|---|
| | Coef.[a] | EI[b] | Coef.[a] | EI[b] | Coef.[a] | EI[b] | Coef.[a] | EI[b] | Coef.[a] | EI[b] |
| NET_D_I | | | 2.842*** | 8.52% | | | | | 2.567*** | 7.67% |
| | | | (3.25) | | | | | | (3.06) | |
| VAL_I | | | | | -0.623*** | -14.67% | | | -0.618*** | -14.56% |
| | | | | | (-6.37) | | | | (-6.32) | |
| INV_I | | | | | | | -0.071*** | -6.59% | -0.056*** | -5.23% |
| | | | | | | | (-4.01) | | (-2.90) | |
| VOLP | 1.984*** | 9.97% | 1.858*** | 9.31% | 2.09*** | 10.53% | 1.97*** | 9.89% | 1.934*** | 9.70% |
| | (5.28) | | (4.77) | | (5.71) | | (5.24) | | (5.06) | |
| Debt_Cost | 0.706*** | 8.62% | 0.669*** | 8.15% | 0.588*** | 7.13% | 0.706*** | 8.62% | 0.56*** | 6.78% |
| | (3.33) | | (3.18) | | (2.78) | | (3.33) | | (2.65) | |
| EP | -0.155*** | -4.30% | -0.155*** | -4.30% | -0.119*** | -3.32% | -0.156*** | -4.32% | -0.121*** | -3.37% |
| | (-3.36) | | (-3.40) | | (-2.76) | | (-3.40) | | (-2.87) | |
| SIZE | -0.114*** | -10.71% | -0.12*** | -11.24% | -0.095*** | -9.01% | -0.114*** | -10.71% | -0.103*** | -9.73% |
| | (-3.37) | | (-3.51) | | (-2.84) | | (-3.39) | | (-3.03) | |
| cons | -0.065 | | -0.052 | | -0.159 | | -0.054 | | -0.131 | |
| | (-0.34) | | (-0.27) | | (-0.83) | | (-0.28) | | (-0.68) | |
| Time Effect | YES | | YES | | YES | | YES | | YES | |
| Industry Effect | YES | | YES | | YES | | YES | | YES | |
| number of group | 73 | | 73 | | 73 | | 73 | | 73 | |
| number of obs | 3212 | | 3212 | | 3212 | | 3212 | | 3212 | |
| Wald Test null: all coefficients=0 | 0[†] | | 0[†] | | 0[†] | | 0[†] | | 0[†] | |
| Pseudo-$R^2$ | 0.3951 | | 0.3976 | | 0.4105 | | 0.3965 | | 0.4145 | |

*Notes:* Estimation results are obtained by using GMM with instrumental variables of the baseline balanced Poisson panel regressions, see Eq. (10a). The dependent variable is the conditional coexceedance of the real economy industries. Independent variables are defined in Table 3. The sample spans from 1Q2001 to 4Q2011. All independent variables are lagged one quarter. The numbers in brackets are t-statistics adjusted for clustering at the industry level.

[a] The estimated coefficient of regressions.
[b] Economic impact, which is computed as detailed in Footnote 20 of our main context.
[†] The p-value of the Wald test.
* Significant level at 10%.
** Significant level at 5%.
*** Significant level at 1%.

**Table 8**
CCX: The impact of industry characteristics in crisis and non-crisis periods.

|  | Model 1 | | Model 2 | | Model 3 | |
|---|---|---|---|---|---|---|
|  | Coef.[a] | EI[b] | Coef.[a] | EI[b] | Coef.[a] | EI[b] |
| NET_D_I*non-crisis dummy | 1.855 * | 4.87% |  |  |  |  |
|  | (1.91) |  |  |  |  |  |
| NET_D_I*crisis dummy | 3.066 ** | 4.46% |  |  |  |  |
|  | (2.50) |  |  |  |  |  |
| NET_D_I |  |  | 2.54 *** | 7.58% | 2.571 *** | 7.68% |
|  |  |  | (3.06) |  | (3.07) |  |
| VAL_I*non-crisis dummy |  |  | -0.572 *** | -11.84% |  |  |
|  |  |  | (-5.77) |  |  |  |
| VAL_I*crisis dummy |  |  | -0.67 *** | -7.99% |  |  |
|  |  |  | (-3.94) |  |  |  |
| VAL_I | -0.614 *** | -14.48% |  |  | -0.62 *** | -14.61% |
|  | (-6.26) |  |  |  | (-6.34) |  |
| INV_I*non-crisis dummy |  |  |  |  | -0.079 ** | -5.48% |
|  |  |  |  |  | (-2.36) |  |
| INV_I*crisis dummy |  |  |  |  | -0.049 ** | -3.10% |
|  |  |  |  |  | (-2.11) |  |
| INV_I | -0.056 *** | -5.23% | -0.055 *** | -5.14% |  |  |
|  | (-2.86) |  | (-2.76) |  |  |  |
| VOLP | 1.922 *** | 9.64% | 1.905 *** | 9.55% | 1.94 *** | 9.74% |
|  | (4.99) |  | (4.74) |  | (5.07) |  |
| Debt_Cost | 0.561 *** | 6.79% | 0.569 *** | 6.89% | 0.558 *** | 6.75% |
|  | (2.66) |  | (2.70) |  | (2.64) |  |
| EP | -0.12 *** | -3.34% | -0.118 *** | -3.29% | -0.121 *** | -3.37% |
|  | (-2.85) |  | (-2.77) |  | (-2.86) |  |
| SIZE | -0.102 *** | -9.64% | -0.103 *** | -9.73% | -0.103 *** | -9.73% |
|  | (-3.01) |  | (-3.05) |  | (-3.03) |  |
| cons | -0.123 |  | -0.125 |  | -0.127 |  |
|  | (-0.64) |  | (-0.66) |  | (-0.66) |  |
| Time Effect | YES |  | YES |  | YES |  |
| Industry Effect | YES |  | YES |  | YES |  |
| number of group | 73 |  | 73 |  | 73 |  |
| number of obs | 3212 |  | 3212 |  | 3212 |  |
| Wald Test null: all coefficients=0 | 0[†] |  | 0[†] |  | 0[†] |  |
| Pseudo-$R^2$ | 0.4178 |  | 0.4229 |  | 0.4148 |  |

*Notes:* Estimation results are obtained by using GMM with instrumental variables of the baseline balanced Poisson panel regressions as shown in equation (10b). The dependent variable is the conditional coexceedance of the real economy industries. Independent variables are defined in Table 3. The sample spans from 1Q2001 to 4Q2011. We split our three main industry variables into two variables: the first variable represents industry variable times non-crisis dummy which equals to one before the second quarter of 2007 and after the first quarter of 2009, otherwise zero, and the second variable represents industry variable times crisis dummy which equals to one between the second quarter of 2007 and the first quarter of 2009, otherwise zero. The numbers in brackets are t-statistics adjusted for clustering at the industry level.

[a] The estimated coefficient of regressions.
[b] Economic impact, which is computed as detailed in Footnote 20 in our main context.
[†] The p-value of the Wald test.
* Significant level at 10%.
** Significant level at 5%.
*** Significant level at 1%.

**Table 9**
CCX: The impact of industry characteristics in competitive and concentrated industries

| | Model 1 | | Model 2 | | Model 3 | | Model 4 | |
|---|---|---|---|---|---|---|---|---|
| | Coef.[a] | EI[b] | Coef.[a] | EI[b] | Coef.[a] | EI[b] | Coef.[a] | EI[b] |
| Panel A: estimation results for competitive industries (number of observations=792) | | | | | | | | |
| NET_D_I*non-crisis dummy | | | 2.765 | | | | | |
| | | | (1.09) | | | | | |
| NET_D_I*crisis dummy | | | 6.085* | 6.53% | | | | |
| | | | (1.78) | | | | | |
| NET_D_I | 4.798** | 12.63% | | | 4.768* | 12.54% | 4.748* | 12.49% |
| | (2.00) | | | | (1.97) | | (1.99) | |
| VAL_I*non-crisis dummy | | | | | -0.812*** | -15.04% | | |
| | | | | | (-4.42) | | | |
| VAL_I*crisis dummy | | | | | -0.834*** | -7.66% | | |
| | | | | | (-2.68) | | | |
| VAL_I | -0.82*** | -16.74% | -0.806*** | -16.48% | | | -0.82*** | -16.74% |
| | (-4.81) | | (-4.74) | | | | (-4.82) | |
| INV_I*non-crisis dummy | | | | | | | -0.152*** | -11.22% |
| | | | | | | | (-2.84) | |
| INV_I*crisis dummy | | | | | | | -0.052 | |
| | | | | | | | (-0.32) | |
| INV_I | -0.112 | | -0.113 | | -0.112 | | | |
| | (-1.61) | | (-1.64) | | (-1.61) | | | |
| Pseudo $R^2$ | 0.438 | | 0.448 | | 0.457 | | 0.438 | |
| Panel B: estimation results for concentrated industries (number of observations=792) | | | | | | | | |
| NET_D_I*non-crisis dummy | | | 0.013 | | | | | |
| | | | (0.01) | | | | | |
| NET_D_I*crisis dummy | | | 2.531 | | | | | |
| | | | (1.16) | | | | | |
| NET_D_I | 1.867 | | | | 1.886 | | 1.848 | |
| | (1.08) | | | | (1.09) | | (1.07) | |
| VAL_I*non-crisis dummy | | | | | -0.580*** | -12.45% | | |
| | | | | | (-3.12) | | | |
| VAL_I*crisis dummy | | | | | -0.739* | -10.72% | | |
| | | | | | (-1.90) | | | |
| VAL_I | -0.674*** | -17.14% | -0.672*** | -17.09% | | | -0.668*** | -17.00% |
| | (-2.74) | | (-2.70) | | | | (-2.64) | |
| INV_I*non-crisis dummy | | | | | | | 0.079 | |
| | | | | | | | (0.50) | |
| INV_I*crisis dummy | | | | | | | 0.032 | |
| | | | | | | | (0.33) | |
| INV_I | 0.042 | | 0.045 | | 0.052 | | | |
| | (0.50) | | (0.53) | | (0.50) | | | |
| Pseudo $R^2$ | 0.429 | | 0.431 | | 0.441 | | 0.431 | |

*Notes:* Estimation results are obtained by using GMM with instrumental variables of the baseline balanced Poisson panel regressions as shown in equation (10a) and (10b). The dependent variable is the conditional coexceedance of the real economy industries. Independent variables are defined in Table 3. The sample contains only competitive industries, from 1Q2001 to 4Q2011. We split our three main industry variables into two variables: the first variable represents industry variable times non-crisis dummy which equals to one before the second quarter of 2007 and after the first quarter of 2009, otherwise zero, and the second variable represents industry variable times crisis dummy which equals to one between the second quarter of 2007 and the first quarter of 2009, otherwise zero. The numbers in brackets are t-statistics adjusted for clustering at the industry level. All models include control variables as used in Table 7 and 8, and industry and time dummy. To save space, we do not report these results here.

[a] The estimated coefficient of regressions.
[b] Economic impact, which is computed as detailed in Footnote 20 in our main context.
† The p-value of the Wald test.
* Significant level at 10%.
** Significant level at 5%.
*** Significant level at 1%.

**Table 10**
Tail risk spillover using distance-to-default.

| Variables | Model 1 | Model 2 |
|---|---|---|
| $DD_{fin}^a$*non-crisis dummy | | 1.508 *** |
| | | (13.71) |
| $DD_{fina}$* crisis dummy | | 1.693 *** |
| | | (14.08) |
| $DD_{fin}^a$ | 1.508 *** | |
| | (13.71) | |
| VOLP | -0.227 | -0.227 |
| | (-1.14) | (-1.14) |
| LEV | -4.671 *** | -4.671 *** |
| | (-17.35) | (-17.35) |
| EP | 0.066 ** | 0.066 ** |
| | (2.23) | (2.23) |
| SIZE | 0.098 *** | 0.098 *** |
| | (3.28) | (3.28) |
| Time Effect | YES | YES |
| Industry Effect | YES | YES |
| Number of groups | 73 | 73 |
| Number of observations | 3139 | 3139 |
| R-squared | 0.6238 | 0.6238 |

*Notes:* Estimation results the baseline balanced panel regressions. The dependent variable is the distance-to-default values for 73 industries. Our sample spans from 2Q2001 to 4Q2011. We estimate the coefficients by means of a Prais-Winsten robust to heteroskedasticity, contemporaneous correlation across panels. The numbers first report the estimated coefficients and below are the t-statistics in brackets adjusted for clustering at the industry level.

[a] The distance-to-default value of financial sector

** Significant level at 5%.

*** Significant level at 1%.